\documentclass[logo]{nvidiatechreport}

\usepackage{graphicx} 
\usepackage[square,numbers]{natbib}

\title{HealDA: Highlighting the importance of initial errors in end-to-end AI weather forecasts}

\usepackage{authblk}

\author[1,*]{Aayush Gupta}
\author[1]{Akshay Subramaniam}
\author[1]{Michael S. Pritchard}
\author[1]{Karthik Kashinath}
\author[2]{Chong-Chi Tong}
\author[2]{Sergey Frolov}
\author[3]{Kelsey Lieberman}
\author[3]{Christopher Miller}
\author[3]{Nicholas Silverman}
\author[1,*]{Noah D. Brenowitz}

\affil[1]{NVIDIA Corporation}
\affil[2]{NOAA}
\affil[3]{MITRE Corporation}
\affil[*]{Corresponding authors: Aayush Gupta (\url{aaygupta@nvidia.com}); Noah D. Brenowitz (\url{nbrenowitz@nvidia.com})}

\date{\today}

\usepackage{tikz}
\usepackage[disable]{todonotes}
\usetikzlibrary{arrows.meta, positioning, shapes}
\usepackage{hyperref}
\usepackage{pgfgantt}
\usepackage[ruled,vlined]{algorithm2e}
\usepackage{amsmath, amssymb}
\usepackage{enumitem}
\usepackage{siunitx}
\usepackage{xcolor}
\usepackage{booktabs}
\usepackage{subcaption}
\usepackage{multirow}

\usepackage{lineno}

\newcommand{\comment}[1]{}

\begin{abstract}
Machine-learning (ML) weather models now rival leading numerical weather prediction (NWP) systems in medium-range skill. However, almost all still rely on NWP data assimilation (DA) to provide initial conditions, tying them to expensive infrastructure and limiting the practical speed and accuracy gains of ML. More recently, ML-based DA systems have been proposed, which are often trained and evaluated end-to-end with a forecast model, making it difficult to assess the quality of their analysis fields. We introduce \emph{HealDA}, a global ML-based DA system that maps a short window of satellite and conventional observations directly to a 1° atmospheric state on the Hierarchical Equal Area isoLatitude Pixelation (HEALPix) grid, using a smaller sensor suite than operational NWP. We treat HealDA strictly as a DA module: its analyses are used to initialize off-the-shelf ML forecast models without any fine-tuning of either. For a variety of off-the-shelf ML forecast models, including FourCastNet3 (FCN3), Aurora, and FengWu, HealDA-initialized forecasts lose less than one day of effective lead time when scored against ERA5. HealDA-initialized FCN3 ensembles similarly trail those of the ECMWF Integrated Forecasting System Ensemble (IFS ENS) system by less than 24~h. We find that forecast error growth in these models is largely unchanged from HealDA initialization, and the skill gap primarily arises from the larger initial-condition error of the HealDA analysis. Spectral analysis reveals that this stems from overfitting to the large scales and upper-tropospheric fields. We also demonstrate that small changes in the verification setup can shift apparent skill by 12--24~h, underscoring the need for consistent scoring. Taken together, these results clarify the current performance of ML-based DA systems and show that a relatively simple, direct observation-to-state network can already provide initial conditions that are usable by state-of-the-art ML forecast models with only modest loss in medium-range skill.
\end{abstract}
\begin{document}

\maketitle

\abscontent

\section{Introduction}

Because the weather is chaotic, the error in a forecast at a given lead time $t$ is fundamentally limited by error in the initial condition \citep{Lorenz1963-hv}.
Mathematically, this can be expressed
\[||\mathbf{\delta x}_t|| \approx ||\mathbf{\delta  x}_0|| e^{(\lambda + \delta \lambda) t}\]
where $\delta x_0$ is the deviation of the initial condition from the true state of the atmosphere,  $\lambda>0$ is the dominant Lyapunov exponent of the weather, and $\delta \lambda>0$ is the increase in this growth rate due to model error.
In the atmosphere, the value of $\lambda$ corresponds to forecast error roughly doubling per day.
This decomposition suggests two credible paths towards improving forecast error: reducing $\delta \lambda$ or $\delta x_0$.

So far, the most popular approach in ML weather science has been to reduce the error growth rate $\lambda + \delta \lambda$, either by authentically reducing model errors to be smaller $\delta \lambda >0$, or artificially reducing variance through machine learning-induced dissipation $\delta \lambda < 0$.
The first method yields real improvements in models by directly training on reanalysis datasets \citep{lam2023learning,bi2022pangu}, while the second artificially reduces deterministic error metrics by suppressing high-frequency details and in so doing, sacrificing ensemble calibration and physical realism.
Such ad-hoc dissipation techniques in ML can include seemingly reasonable ideas like training deterministic models on long lead time \citep{Nguyen2023-sm}, multi-step fine tuning \citep{Brenowitz2018-td,Weyn2019-rr,lam2023learning}, and replay buffers \citep{FengWu}, but all these approaches act to make individual forecasts behave like the ensemble mean.
Probabilistic benchmarks point out the shortcomings of these variance reduction techniques \citep{BrenowitzLaggedEnsemble,Rasp2024-yl}, 
and despite these early missteps, the current generation of ML models is mostly probabilistic now \citep{price2023gencast,NeuralGCM}, and it is clear that $\delta \lambda$ is now smaller than it once was.

However, with each additional advance in reanalysis-trained models, further improvements to model error $\delta \lambda$ become more difficult, so it is essential to reduce $||\delta x_0||$ further through better data assimilation systems.
In operational NWP, data assimilation is the glue that connects the observing system and the forecast model. Every 6 hours, millions of heterogeneous observations (satellite radiances, aircraft, radiosondes, surface networks, radar) are combined with a short-range background forecast to produce a best-estimate analysis state \citep{ecmwf_obs}. Modern 4D-Var \citep{4dvar} and ensemble-variational systems solve a high-dimensional optimization problem at each cycle, iterating the forecast model and its adjoint many times, making data assimilation a major computational and operational bottleneck: the European Centre for Medium-Range Weather Forecasts (ECMWF) reports that DA---including the high-resolution analyses and the ensemble of data assimilations (EDA)---consumes $\sim 5\times10^3$ CPU node-hours per day, roughly $\sim$40\% of the compute in their operational breakdown \citep{buizza2018tm829}. As a result, even if the forecast step becomes cheap using ML weather models, continued reliance on NWP DA means that the overall system inherits the latency, cost, and complexity of the upstream NWP DA pipeline, which typically operates on fixed 6-hour cycling schedules. These DA systems are also heavily hand-tuned: adding or updating new sensors involves updating error models, biases, and quality control rules through substantial calibration and slow operational tuning. Thus, an attractive alternative is to replace not only the forecast model but also the complex DA system with learned components that ingest raw observations directly.

While it might seem obvious that a DA system should seek to minimize initial-condition error to lie within the uncertainty of modern reanalyses, a scan of the recent ML DA literature (\citep{Allen2025EndToEndWeather, ni2025huracan, wang2025xicheno, Sun2025Fuxi}) and our results below show that current ML-based DA systems, including our own, still exhibit errors that are roughly twice what we estimate to be the error of ERA5 with respect to the true atmospheric state, based on proxies such as inter-analysis differences and short-range forecast behavior. We speculate this fact has been overlooked in the still nascent ML DA literature due to 1) a focus on end-to-end performance and 2) different verification standards. End-to-end forecast error is often used as a proxy for the skill of DA systems simply because it is easier to statistically detect an error once it has grown by a factor of $e^{(\lambda + \delta\lambda) t}$.
This method of verification is reasonable, but in the ML context, it has introduced an array of confounding factors related to the error growth of the forecast model ($\delta \lambda$) that makes comparison difficult across ML DA systems. For example, many works only assess deterministic skill scores, which are susceptible to unphysical variance reduction methods like multi-step fine-tuning (see \citep{Allen2025EndToEndWeather}), or simply build a strong probabilistic forecast model \citep{ni2025huracan}. In these cases, it is likely that the presented forecast model would perform even better on ERA5 initial conditions, which still leaves the DA problem unsolved.

Another issue has been that subtleties in the verification procedure can easily shift skill curves by 12-24 hours, which is the magnitude required to achieve an apparently state-of-the-art result. By contrast, typical differences in forecast skill between leading operational centers are much smaller, with 24-hour improvement equivalent to a decade of traditional model development, underscoring the need for careful and consistent verification procedures. For example, \citet{ni2025huracan} incorrectly assume that the Continuous Ranked Probability Score (CRPS) remains a proper score when computed against different ground truth datasets. Section \ref{sec:scoring-choice} shows this can shift scores by a day of accuracy in favor of the forecast that happens to be scored against a blurrier reference analysis, which is usually the ML model. Similarly, differing observation look-aheads of various analysis products can shift forecast error curves by 12 hours or more of accuracy \citep{lam2023learning}. As a result of these factors, it is difficult to determine whether a given ML DA system is actually producing analysis fields that can stand in for ERA5 or IFS analyses—i.e., generic, truth-like initial conditions that any forecast model could use with low error ($\delta x_0$)—or if the reported skill may mostly reflect tweaks to error growth ($\delta \lambda$) and a flawed verification process biased toward optimistic results.

To help disentangle the differing effects of error growth $\lambda$ and initial error $\delta x_0$, we introduce \emph{HealDA}, a global observation-to-state ML DA system, trained independently of any forecast model and not provided a gridded background state as a model input field at runtime. 
We show that HealDA analyses are nearly as informative as ERA5 for modern probabilistic and deterministic ML forecast models: when used to initialize off-the-shelf systems such as FCN3 \citep{bonev2025fcn3}, Microsoft's Aurora \citep{bodnar2025aurora}, or FengWu \citep{FengWu}, forecast error growth closely tracks ERA5-initialized runs, with a loss of less than one day of effective lead time. This demonstrates that the DA and forecast problems can be cleanly decoupled, and that, once analysis error is sufficiently small, ML-based DA can supply initial conditions that are effectively interchangeable with ERA5-like analyses for state-of-the-art ML forecast models. To our knowledge, this is the first demonstration that ML-based DA initial conditions can be used in this plug-and-play way with existing high-skill ML forecast models.

We further explain these results through a spectral analysis of error growth and show that the forecast error growth ($\lambda$) is not increased significantly by using HealDA, and make a key observation that \emph{the increased analysis error on the DA task when using ML is caused by overfitting to the largest scale modes of the atmosphere}. Together with our simple architecture, we hope that these insights provide a platform and productive direction for future research into ML DA methods.

\section{Related Work}

\paragraph{Forecasts model trained on reanalysis} Early work in ML-based weather models focused solely on the forecasting task while still relying on traditional Numerical Weather Prediction (NWP) based data assimilation (DA) for the initial conditions. FourCastNet \citep{pathak2022fourcastnet} demonstrated that ML weather models trained on ERA5 reanalysis data could produce global deterministic forecasts at 0.25° resolution out to 10 days. Models like GraphCast \citep{lam2023learning}, Pangu-Weather \citep{bi2022pangu}, and FengWu \citep{chen2023fengwupushingskillfulglobal} further improved deterministic skill, beating even ECMWF's High-Resolution Integrated Forecast System (IFS HRES) on most variables and lead times. More recent work with GenCast \citep{price2023gencast} and FourCastNet3 (FCN3) \citep{bonev2025fcn3} has extended these ideas to probabilistic forecasting to generate ensembles outperforming ECMWF's ensemble prediction system (ENS), all while being almost an order of magnitude faster. All of these models, however, assume access to high-quality analyses to initialize the forecast (e.g., ERA5, IFS analysis) and are therefore still highly dependent on traditional DA systems.

\paragraph{ML Data assimilation} Several recent ML-based systems take a step towards skillful end-to-end forecasts driven from raw observations. Aardvark Weather \citep{Allen2025EndToEndWeather} processes heterogeneous raw observations to produce 10-day global forecasts at 1.5° resolution, and shows that a purely observational ML model can approach operational NWP skill, but it only matches or exceeds IFS HRES at the longest lead times, where its fields are visibly smoothed with the known blurring effects of multi-step fine-tuning in deterministic models such as GraphCast\citep{BrenowitzLaggedEnsemble}. FuXi Weather \citep{Sun2025Fuxi} adopts a similar idea of DA initialization followed by a forecast model, and finds that including a short-range background forecast is crucial for stabilizing the DA problem under sparse observations. Its cycling system extends the Z500 skillful horizon beyond HRES, but again shows the gains only at later lead times with heavily smoothed forecasts. Further, its analysis error against ERA5 remains high, generally comparable to or even higher than that of a 42-hour FuXi forecast. In contrast, XiChen \citep{wang2025xicheno} adopts a 4D-Var-like method, but approaches similar skill and deterministic forecast quality to Aardvark and FuXi. Huracan \citep{ni2025huracan} tackles the same observation-to-forecast problem in the probabilistic regime: a stochastic DA model and a stochastic forecast model are trained to produce 1° ensemble forecasts. Their reported CRPS values meet or exceed ECMWF ENS for most variable–lead combinations \emph{when each system is verified against its own analysis}, with substantial gains for temperature and humidity fields but a remaining gap for the geopotential fields. They do not show results verified against ERA5, which we show in Section \ref{sec:scoring-choice} results in artificially good CRPS scores due to the use of a blurry verification dataset.

\paragraph{Generative score-based DA} A separate line of work uses diffusion models as generative priors for DA. Score-based Data Assimilation (SDA) \citep{rozet2023scorebaseddataassimilation} and its regional extension to km-scale weather \citep{manshausen2025generativedataassimilationsparse} demonstrate that sparse real-world surface observations can guide diffusion models to produce analysis-like fields, while methods such as DiffDA, Appa, and LO-SDA \citep{huang2024diffda,andry2025appabendingweatherdynamics,sun2025losda} extend guided diffusion DA to global settings. At present, however, these global approaches have only been tested in idealized experiments with synthetic observation setups. This slow progress is likely because SDA requires two models 1) the raw diffusion model and 2) the observation operator. For conventional observations, the observation operator is trivial---just interpolation---but generating realistic satellite irradiance from physical state space is more difficult and error prone. Moreover, SDA involves numerous approximations which introduce both error and new hyper-parameters for the relative weight of the prior and each observation stream that must be carefully tuned. This all makes SDA, and inverse-problem approaches more broadly, less immediately applicable for the global forecast problem, where the observing system is considerably more heterogeneous and the baselines are more accurate.

\paragraph{Forecasting in observation space} A complementary direction avoids gridded analyses altogether and operates directly in observation space. Frameworks such as GraphDOP, AI-DOP, and DAWP \citep{mcnally2024dop, mcnally2025aidop, gong2025dawp} train models to forecast future observations rather than gridded state variables, thereby sidestepping reanalysis biases during training. However, such results are still preliminary, and so far, have achieved useful skill primarily at shorter lead times.

\section{HealDA}

\begin{figure}
    \centering
    \includegraphics[width=1\linewidth]{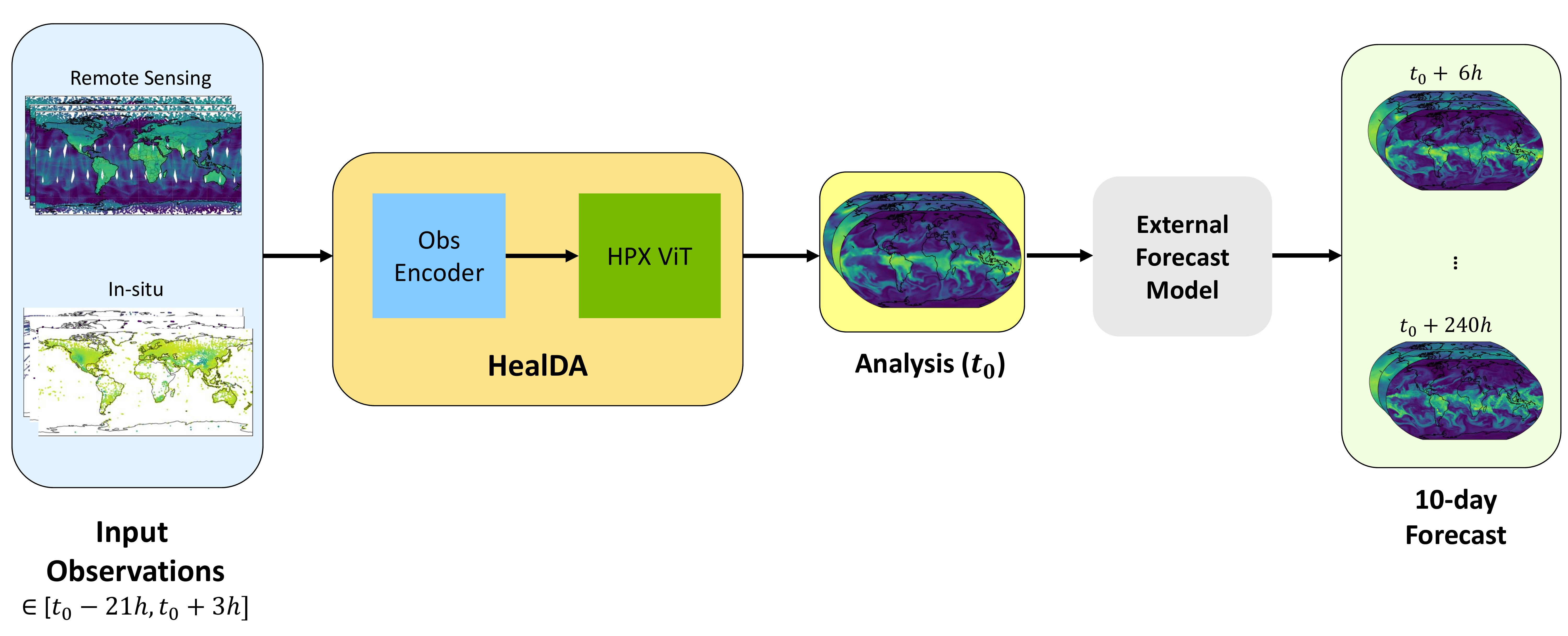}
    \caption{\textbf{End-to-end HealDA system and forecasting pipeline.} Observations from various remote-sensing instruments (ATMS, MHS, etc.) and in-situ sources (radiosondes, buoys, etc.) in the time window $[t_0 - 21\,\mathrm{h},\, t_0 + 3\,\mathrm{h}]$ are processed by HealDA, which consists of an Observation Encoder (Obs Encoder) followed by an HPX ViT backbone, to produce an analysis state on the HPX grid at the target time $t_0$. This analysis can then serve as the initial condition for an external forecast model (e.g., FCN3), to produce a 10-day forecast $(t_0 + 6\,\mathrm{h}, \dots, t_0 + 240\,\mathrm{h})$.}
    \label{fig:system-diagram}
\end{figure}

HealDA is a global ML-based DA system trained to map a short window of observations to the ERA5 reanalysis states on the $1^\circ$ HPX64 grid, corresponding to $N_{pix}=49{,}152$ equal-area HEALPix pixels (see Section~\ref{subsec:healpix} for grid details). Given a 24-hour window of satellite and conventional observations from $t-21$~h to $t+3$~h around the target analysis time $t_0$, the model produces a 74-channel state: five atmospheric variables (temperature, specific humidity, geopotential, and zonal/meridional wind) at 13 pressure levels, along with nine surface variables (Table~\ref{tab:healda-outputs}).

HealDA ingests heterogeneous observational data from microwave sounders, GNSS radio occultation, in-situ networks (e.g., radiosondes, aircraft, buoys), and satellite-derived wind products. Observations are ingested as an unordered point cloud, where each scalar measurement is represented together with its associated continuous metadata—including observation spatial position, time, viewing geometry, pressure, and height—as well as discrete identifiers such as sensor type, satellite platform, and sensor channel. Across the 24-hour window, HealDA ingests on the order of $10M$ individual scalar observations globally, with strong spatial and temporal heterogeneity depending on the observing system.

HealDA consists of two main components: an observation encoder followed by an HPX vision transformer (ViT) backbone (Figure~\ref{fig:healda-arch}). The observation encoder processes heterogeneous, point-cloud-like observational data by treating each sensor type (e.g., AMSU-A, MHS, ATMS, conventional) with a dedicated encoder (Figure ~\ref{fig:healda-arch}). The obs encoder operates directly on individual observations with continuous time embeddings and aggregates only after tokenization. In comparison, existing ML DA approaches typically handle observation windows through pre-gridding and early aggregation, either by retaining only the most recent observation per grid cell (e.g.,~\citep{Allen2025EndToEndWeather}) or by discretizing the observation window into a small number of fixed temporal bins that are encoded as separate channels (e.g.,~\citep{ni2025huracan, Sun2025Fuxi}).

Observations are embedded together with their associated metadata and aggregated onto the HPX64 grid using scatter--reduce operations, producing per-sensor gridded feature maps. These are fused into a single dense representation and passed to the HPX ViT backbone, which globally integrates sparse observational embeddings to infer the complete atmospheric state, outputting the analysis at the target time. The ViT uses diffusion transformer ( DiT)-style blocks and is adapted to the HEALPix grid, with patch-based encoding and decoding and global self-attention across the sphere. Both components are trained jointly end-to-end under a single supervised regression objective. See Section \ref{sec:methods} for full training and architectural details.

Importantly, HealDA is trained as a standalone observation-to-state ML DA module without a gridded background state provided as a model input field, using ERA5 as supervision, and independent of any forecast model. Below, we show that the resulting analyses can be used as plug-and-play initial conditions for off-the-shelf forecast models such as FCN3, Aurora, and FengWu, which we keep fixed and do not fine-tune. When coupled with a probabilistic forecast model such as FCN3, HealDA yields an ensemble forecasting system driven by observational inputs. Inference is computationally lightweight: on a single NVIDIA H100 GPU, HealDA can produce a global analysis in under one second. This analysis can then be used to initialize FCN3, which can produce a single-member, 10-day forecast at 6-hourly resolution in under one minute on a single H100 GPU. In contrast, ECMWF's EDA analyses used to initialize IFS ENS require $\sim$1\,h wall-clock time on $\sim$1800 CPU nodes \citep[Table~2]{buizza2018tm829}. Our end-to-end forecasting pipeline is illustrated in Figure~\ref{fig:system-diagram}.

\section{Results}

HealDA is a global DA system that produces a best guess of the atmospheric state in an observation-to-state framework. We evaluate HealDA in terms of both the quality of its analysis and its impact on downstream ML forecasting models. We compare the error of HealDA analysis and operational IFS analysis against ERA5. To investigate how interchangeable HealDA analyses are with ERA5 for downstream ML forecast models, we initialize off-the-shelf probabilistic (FCN3) and deterministic (Aurora, FengWu) networks with ERA5 and HealDA initial conditions. For FCN3, we additionally compare against ECMWF’s IFS ENS to place HealDA-initialized ensemble skill in the context of the current operational standard.

We also examine HealDA's robustness to changes in the observing system. We further analyze HealDA analysis and forecasts using spectral diagnostics to identify the primary source of the error. We note that small changes in the verification setup, such as the scoring reference or differing lookaheads, can shift apparent forecast skill by $12\text{ h}$ or more of effective lead time, highlighting the need for consistent scoring when comparing ML-based DA systems. All headline results here use ERA5 as the reference, and a detailed discussion of scoring effects is deferred to the Supplementary Information (see Sections \ref{sec:scoring-choice}, \ref{sec:init-time}).

\subsection{Analysis Fields}
\label{sec:analysis-rmse}

\begin{figure}
    \centering
    \includegraphics[width=1\linewidth]{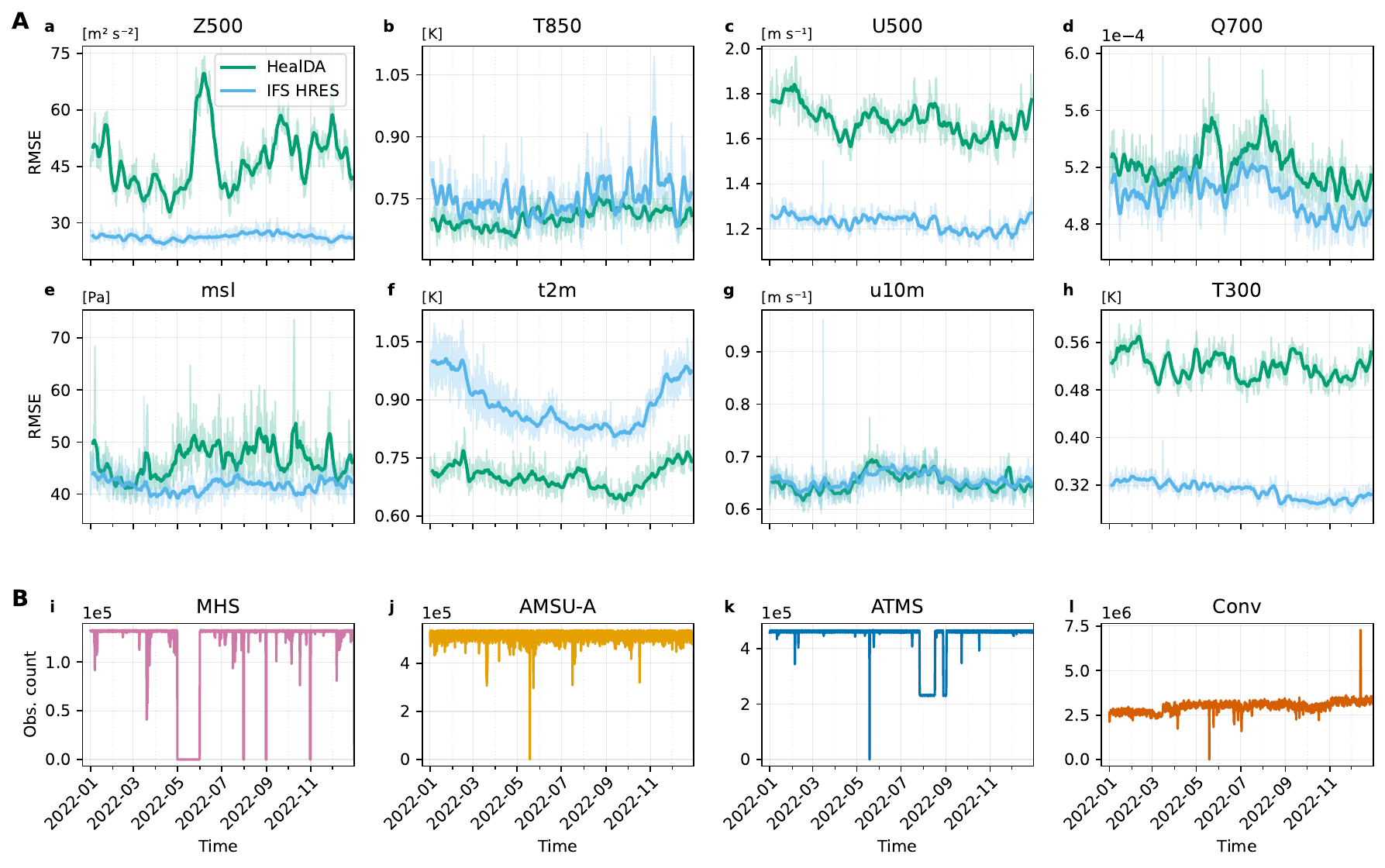}
    \caption{\textbf{RMSE of HealDA analysis vs IFS.} \textbf{(A)} Time series of global RMSE for both HealDA and IFS against ERA5 in the 2022 test period, computed every 6 hours (00/06/12/18 UTC). The original data is shown with reduced opacity to reduce noise, and the solid line represents the 7-day moving average. \textbf{(B)} Corresponding observation counts of 6-hour windows centered at each analysis time for HealDA's sensor suite: (i) MHS, (j) AMSU-A, (k) ATMS microwave sounders, and (l) all conventional observations}
    \label{fig:analysis-rmse}
\end{figure}

To place HealDA’s analysis errors in context, we compare our analyses to the initial conditions ECMWF uses to initialize its forecasts in Figure~\ref{fig:analysis-rmse}, scoring both against ERA5 on the HPX64 grid over the 2022 test period. Specifically, we use the $t=0$ field from the operational IFS HRES system. We note that post–2017 HRES operates with a newer DA cycle with improved model physics relative to that used in ERA5, and is thus better aligned with the true observations, particularly for surface variables \citep{haiden2023forecast}.
We stress that because they share the same input data and have similar components, the ERA5-HRES discrepancy likely underestimates the true error of ERA5---especially where ERA5 and HRES share similar mean state biases.
Nonetheless, since HealDA, like other AI DA models, is explicitly trained to emulate ERA5, it is fair to hope that it should deviate from ERA5 by less than its sibling analysis HRES. Unfortunately, the results below show this is not the case.

For geopotential fields such as Z500, HealDA’s RMSE is typically in the $\sim\!50\,{m^2/s^2}$ range, roughly a factor of two larger than the $\sim\!25\,{m^2/s^2}$ deviation of HRES from ERA5. A similar pattern appears for winds and upper-level temperature fields (e.g., U500 and T300), where HealDA’s deviations from ERA5 are approximately 50--100\% larger than those of HRES. Closer to the surface, this gap steadily shrinks: by 850--1000\,hPa, the RMSE of HealDA and HRES for $T$, $U$, and $V$ becomes much more comparable. Specific humidity fields ($Q$) behave similarly but with overall smaller errors, and the HealDA and HRES Q700 time series track each other closely.

This behavior is broadly consistent with strong observational constraints on temperature and humidity from microwave sounders, together with substantial—but vertically and regionally uneven—constraints on winds from satellite wind retrievals. By contrast, the structure of geopotential must be inferred indirectly from those profiles and dynamical balance, so it is not surprising that HealDA geopotential fields are most sensitive and lag HRES most clearly. A caveat is that at upper atmosphere levels, HealDA exhibits larger deviations from ERA5 across most fields, which we later show is due to overfitting at large spatial scales (see Section \ref{sec:spectral-error}). Note that large RMSE spikes often coincide with observing-system outages (see Section \ref{sec:robustness}).

Near the surface (msl, t2m, and 10\,m winds; see Table \ref{tab:healda-outputs} for variable definitions), HealDA’s deviation from ERA5 at first appears comparable to, and in some fields less than, that of HRES. However, in the variables where HealDA and HRES exhibit similar RMSE, subsequent analysis will reveal that part of HealDA's advantage can be attributed to small-scale smoothing in the HealDA analysis (Sections~\ref{sec:spectral-error}--\ref{sec:spectra}).

\begin{figure}
    \centering
    \includegraphics[width=0.75\linewidth]{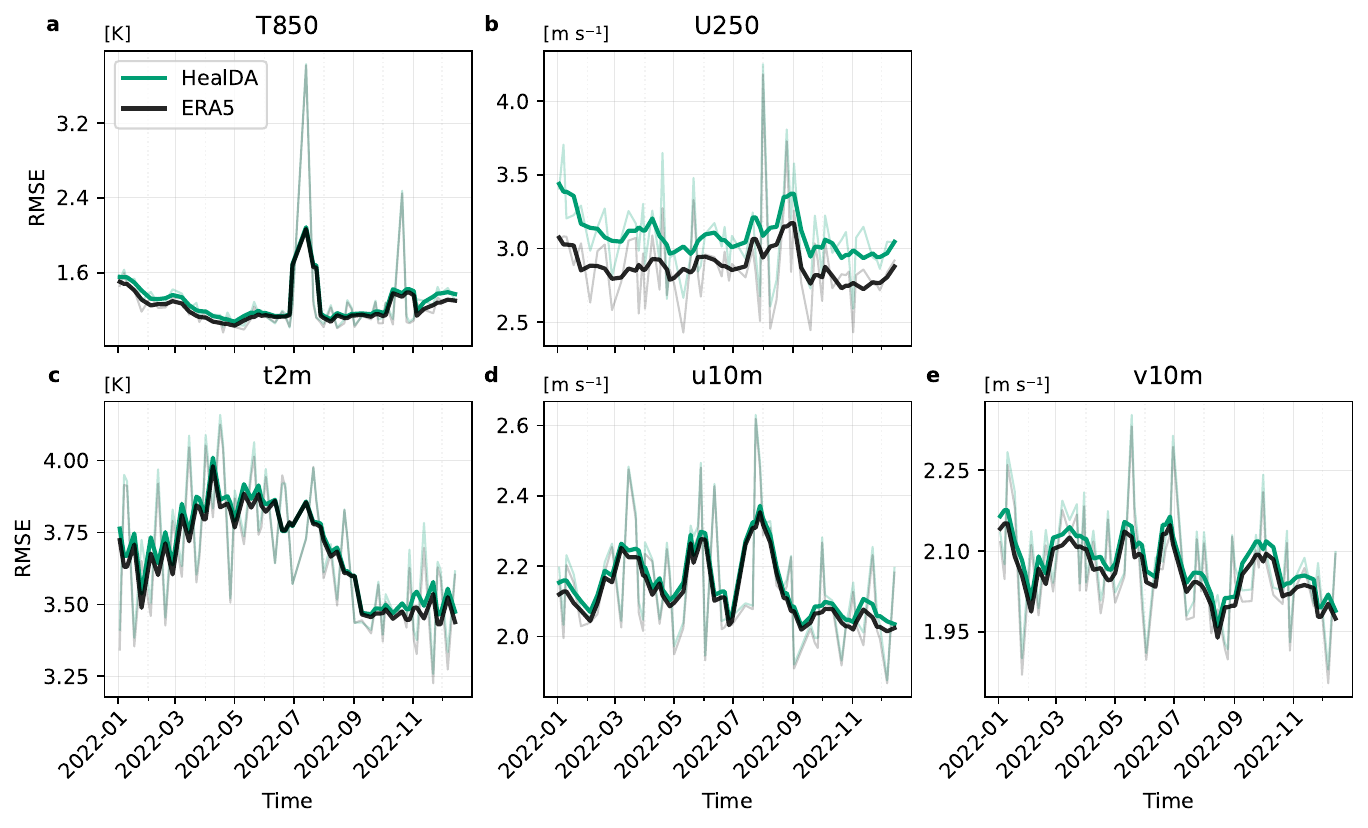}
    \caption{\textbf{Six-hour forecast verification against in-situ observations.} RMSE time series of 6-hour FCN3 forecasts initialized from ERA5 and HealDA analyses and verified against in-situ observations in the 2022 test period using the wxvx workflow. Forecasts are initialized at 06/18 UTC and verified at 00/12 UTC. The original data is shown with reduced opacity to reduce noise, and the solid line represents the 28-day moving average.}
    \label{fig:wxvx-obs-rmse}
\end{figure}

To further assess analysis quality, in Figure \ref{fig:wxvx-obs-rmse}, we evaluate 6-hour forecasts initialized from ERA5 and HealDA analyses against in-situ observations using the wxvx package \citep{wxvx}, which drives the Model Evaluation Tools (MET) \citep{MET}. We use FourCastNet3 to produce an ensemble of $N=16$ forecasts for both analyses. Verification of the ensemble mean is performed against observations occurring in the subsequent 6-hour window, which were not used in the assimilation and can therefore be treated as independent of the assimilation step. Additionally, at this short lead time, the contribution of the model error growth should be minimal, such that the error against observations primarily reflects the analysis quality. We restrict to 06/18 UTC initializations, where ERA5's observation lookahead matches HealDA's, and this corresponds to verification at 00/12 UTC, where observation density is higher relative to 06/18 UTC. HealDA exhibits slightly larger errors than ERA5 across all variables, and it is most pronounced in U250, an upper-tropospheric field. This mirrors the analysis discrepancies seen relative to IFS HRES, reinforcing that HealDA's largest errors are at these upper atmospheric levels.

\subsection{End-to-end Probabilistic Skill}
\begin{figure}
    \centering
    \includegraphics[width=1\linewidth]{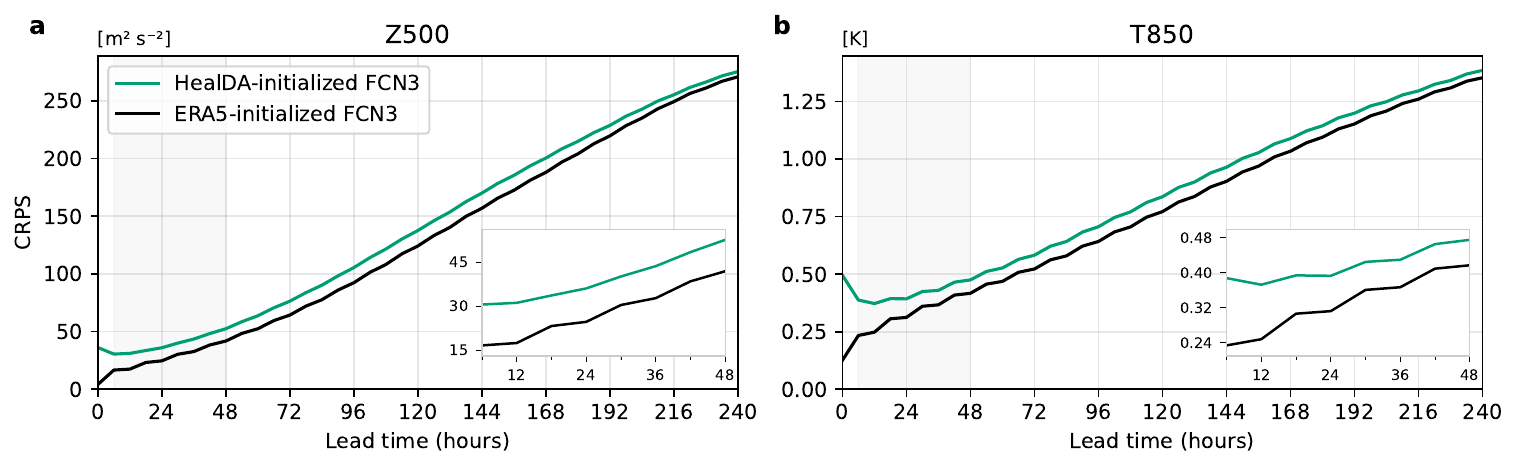}
    \caption{\textbf{Probabilistic FCN3 skill with HealDA and ERA5 initial conditions.} CRPS of FCN3 forecasts initialized by HealDA and ERA5, both verified against ERA5 on the HPX64 grid and averaged over 128 initial conditions at 06/18 UTC in 2022. The inset panels zoom into the 6-48~h lead time range.}
    \label{fig:fcn3-healda-vs-era5}
\end{figure}

To isolate the impact of the HealDA initial condition on forecasting, we run the same FCN3 forecasting model first with HealDA analyses and then again with ERA5 analyses (coarsened to consistent HPX64 resolution), with results shown in Figure \ref{fig:fcn3-healda-vs-era5}. We use a 16-member ensemble for FCN3 forecasts.

As detailed in Section \ref{sec:scoring-procedure}, we score all forecasts against ERA5 after coarsening from their native $0.25^\circ$ resolution to the HPX64 grid. Section~\ref{sec:ic-resolution} shows that initializing forecasts from lower-resolution HPX64 analyses (as opposed to $0.25^\circ$) has only a minor impact on forecast skill for most models, corresponding to a $\le 3\,\mathrm{h}$ shift in effective lead time rather than a systematic change in error growth rate. We use 06/18 UTC in 2022 so that ERA5 and HealDA have the same $\,+3$~h observation look-ahead. The impact of 00/12~UTC initializations, which give ERA5 an additional $\,+6$~h of observation look-ahead, is discussed in Section \ref{sec:init-time}.

An immediately important observation is that across all variables, the HealDA-initialized CRPS curves are nearly parallel to the ERA5-initialized ones, with a roughly constant horizontal shift corresponding to a loss of $\le$ 12-18 hours of effective lead time. In other words, replacing ERA5 with HealDA initial conditions does not noticeably change the FCN3's error growth rate $\lambda$---the forecast skill is as if starting from an 18h forecast. This error growth is further investigated in Section \ref{sec:spectral-error}. Additional RMSE and anomaly correlation coefficient (ACC) curves tell the same story, and just like its intrinsic error growth, FCN3's ensemble spread is unchanged under HealDA initialization (see Section \ref{sec:extended-metrics}). This confirms our working hypothesis that \textit{the main impact of using ML-based initial conditions is shifting the starting error $||\delta x_0||$, not altering the subsequent growth rate}. As an aside, a secondary spike in the CRPS at lead time 0 is expected, given that HealDA produces a deterministic analysis, meaning the CRPS at time 0 reduces to the mean absolute error (MAE).

\begin{figure}
    \centering
    \includegraphics[width=1\linewidth]{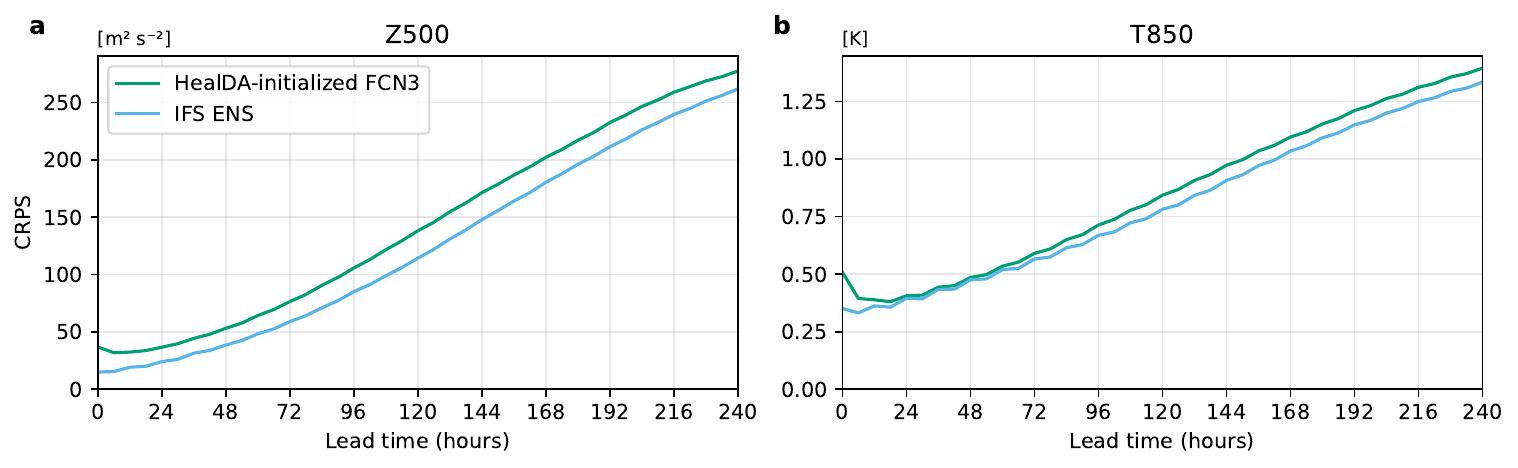}
    \caption{\textbf{Probabilistic skill of HealDA-initialized FCN3 vs IFS ENS.} CRPS of IFS ENS forecasts and FCN3 forecasts initialized from HealDA, verified against ERA5 on the HPX64 grid and averaged over 128 initial conditions at 00/12 UTC in 2022.}
    \label{fig:ifs-comparison}
\end{figure}

Figure \ref{fig:ifs-comparison} compares the probabilistic skill of our HealDA-initialized FCN3 forecasting system to the gold-standard in operational ensemble forecasting---ECMWF's IFS ENS (50 perturbed members). HealDA-initialized FCN3 trails IFS ENS by less than 24h of effective lead time, similar to the previous comparison with ERA5-initialized FCN3. HealDA forecasts tend to have lower spread and poorer spread-skill ratio (SSR) calibration at early lead times (see Figure~\ref{fig:healda-vs-ifs-full}), since HealDA produces a single deterministic analysis, whereas each member of the IFS ENS ensemble is initialized with a perturbed initial condition to capture the uncertainty in the initial condition.

In Table~\ref{tab:da_comparison}, we additionally compare our initial condition and forecast scores to those of prior ML DA approaches, using each method’s reported scores, with IFS ENS included for reference.

\begin{table*}[!t]
\centering
\small
\setlength{\tabcolsep}{5pt}

\begin{subtable}{\textwidth}
\centering
\caption{RMSE Scores (vs ERA5 unless otherwise indicated)}
\label{tab:rmse}
\begin{tabular}{l l ccc ccc ccc}
\toprule
 &  & \multicolumn{3}{c}{IC RMSE} & \multicolumn{3}{c}{RMSE at 1 day} & \multicolumn{3}{c}{RMSE at 5 days} \\
\cmidrule(lr){3-5} \cmidrule(lr){6-8} \cmidrule(lr){9-11}
Method &
Validation Protocol &
Z500 & T850 & U500 &
Z500 & T850 & U500 &
Z500 & T850 & U500 \\
\midrule
ERA5 \citep[Fig. 12]{Hersbach2020ERA5}$^{\ddag}$&
$\sim$ 2018 &
-- & 0.34 & 0.74 &
-- & --& -- &
-- & -- & -- \\
\midrule
 Aardvark~\citep{Allen2025EndToEndWeather}$^{*}$ &
1.5$^\circ$, 2018 &
62 & 0.94 & 2.3 &
93 & 1.0 & 2.8 &
450 & 2.2 & 5.9 \\

FuXi Weather~\citep{Sun2025Fuxi}$^{*}$ &
0.25$^\circ$, 2023--2024 &
80 & 1.0 & 2.2 &
100 & 1.1 & 2.5 &
390 & 2.1 & 5.3 \\

XiChen~\citep{wang2025xicheno}$^{*}$ &
1.40625$^\circ$, 2023 &
100 & 1.1 & -- &
150 & 1.2 & -- &
490 & 2.3 & -- \\

IFS ENS$^\dagger$ &
1$^\circ$, 2022 &
26 & 0.71 & 1.11 &
47 & 0.77 & 1.46 &
280 & 1.62 & 4.22 \\

\multirow{2}{*}{\raisebox{0.6ex}}{HealDA-initialized Atlas~\citep{atlas}$^\dagger$} &
\multirow{2}{*}{HPX64, 2022} &
\multirow{2}{*}{46.6} &
\multirow{2}{*}{0.70} &
\multirow{2}{*}{1.68} &
66.4 & 0.82 & 1.98 &
328 & 1.80 & 4.70 \\

\multirow{2}{*}{\raisebox{0ex}}{HealDA-initialized FCN3$^\dagger$} &
& & & &
67.3 & 0.78 & 1.98 &
326 & 1.76 & 4.71 \\

HealDA (L1)-initialized FCN3$^\dagger$ &
HPX64, 2022 &
49.3 & 0.73 & 1.73 &
70 & 0.81 & 2.03 &
333 & 1.78 & 4.79 \\

\bottomrule
\end{tabular}

\vspace{0.4em}
\raggedright
\footnotesize
$^{\ddag}$ Spread of DA ensemble reported. \\
$^{*}$ Single-member forecast. Trained with multi-step loss functions. \\
$^\dagger$ Ensemble mean scores reported.\\
\end{subtable}

\vspace{1em}

\begin{subtable}{\textwidth}
\centering
\caption{CRPS Scores}
\label{tab:crps}
\begin{tabular}{l c ccc ccc}
\toprule
 &  & \multicolumn{3}{c}{CRPS at 1 day} & \multicolumn{3}{c}{CRPS at 5 days} \\
\cmidrule(lr){3-5} \cmidrule(lr){6-8}
Method &
Validation Protocol &
Z500 & T850 & t2m &
Z500 & T850 & t2m \\
\midrule

IFS ENS &
ERA5, 1$^\circ$, 2022 &
23.9 & 0.38& 0.41 &
115 & 0.77 & 0.61 \\

HealDA-initialized Atlas &
ERA5, HPX64, 2022 &
35.7 & 0.42 & 0.43 &
141 & 0.86 & 0.69 \\

HealDA-initialized FCN3 &
ERA5, HPX64, 2022 &
36.0 & 0.39 & 0.39 &
138 & 0.84 & 0.64 \\

HealDA (L1)-initialized FCN3 &
ERA5, HPX64, 2022 &
38.4 & 0.42 & 0.40 &
142 & 0.86 & 0.65 \\
\midrule
Huracan~\citep{ni2025huracan} &
Analysis, 1$^\circ$, 2024 &
25 & 0.28 & 0.23 &
130 & 0.73 & 0.51 \\

IFS ENS&
Analysis, 1$^\circ$, 2022 &
21.0 & 0.31 & 0.30 &
114 & 0.75 & 0.55 \\

HealDA-initialized Atlas &
Analysis, HPX64, 2022 &
30.1 & 0.35 & 0.34 &
138 & 0.80 & 0.62 \\

HealDA-initialized FCN3 &
Analysis, HPX64, 2022 &
30.5 & 0.34 & 0.30 &
138 & 0.79 & 0.58 \\

HealDA (L1)-initialized FCN3 &
Analysis, HPX64, 2022 &
31.8 & 0.35 & 0.31 &
142 & 0.81 & 0.59 \\
\bottomrule
\end{tabular}

\vspace{0.4em}
\raggedright
\end{subtable}

\vspace{0.4em}
\raggedright
\footnotesize

\caption{Skill comparison of ML DA systems and the operational IFS ENS, showing (a) initial condition (IC) and forecast RMSE with respect to ERA5, and (b) forecast CRPS for probabilistic systems with respect to ERA5 and each method’s analysis. Scores against analysis are included for comparability with Huracan, which reports CRPS only against its analysis. The Validation Protocol column lists the verification spatial resolution and evaluation period in (a), and additionally, the verification reference in (b). Scores for ML DA methods are estimated from their figures using plot digitization software and reported to two significant figures. Our HealDA scores are averaged across 06/18 UTC and 00/12 UTC initialized forecasts. HealDA (L1) scores correspond to HealDA fine-tuned to only take Level~1 products as input (see Section~\ref{sec:lvl2-ablation}). The ERA5 DA ensemble spread is shown as a comparison in (a). IFS ENS scores are reported from WeatherBench2. Units are m$^{2}$\,s$^{-2}$ for Z500, K for T850 and t2m, and m\,s$^{-1}$ for U500.}
\label{tab:da_comparison}

\end{table*}

\subsection{Large-scale Overfitting}
\label{sec:spectral-error}

In this section, we will disaggregate the error in spectral space, noting that error growth in forecasts is dominated by the synoptic scale (\SI{1000}{km}) since this is where baroclinic instability is most active.  In the discussion below, we will use $P(\cdot)$ to denote the spherical power spectra. 

\begin{figure}
    \centering
    \includegraphics[width=1\linewidth]{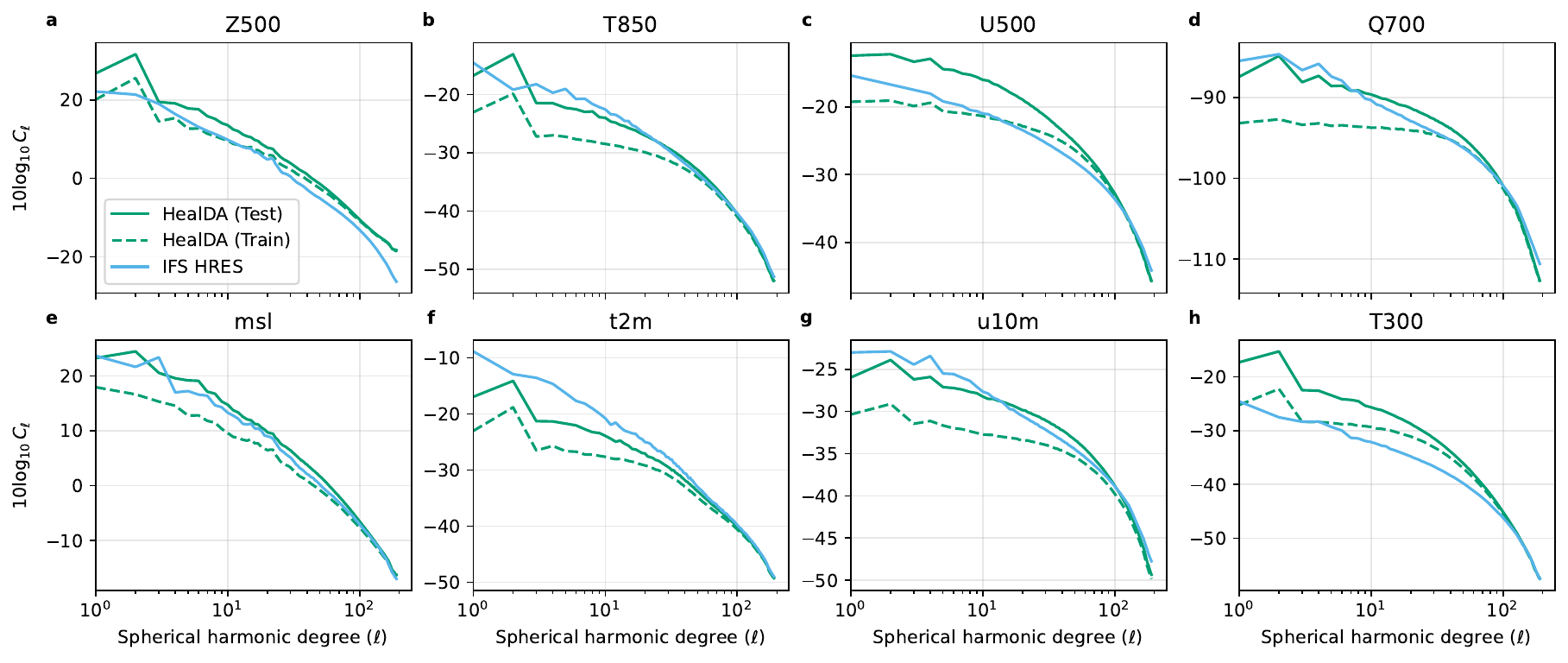}
    \caption{\textbf{Analysis error spectral decomposition.} Spherical power spectra of HealDA and IFS HRES analysis errors on the HPX64 grid, scored relative to ERA5. The HealDA error spectra are shown, averaged over the test year (2022), in solid lines, and a year from the training period (2021), in dashed lines. For IFS, the error spectra averaged across 2021-2022 are shown. Spectra are shown as a function of spherical harmonic degree $\ell$ (large scales on the left, small scales on the right) and plotted as $10\log_{10} C_\ell$, where $C_\ell = \frac{1}{2\ell+1}\sum_{m=-\ell}^{\ell}\left|a_{\ell m}\right|^2$.}
    \label{fig:spectra-t0}
\end{figure}

Figure \ref{fig:spectra-t0} shows the error power spectra of our DA initial conditions versus ERA5, both for the training and test sets. IFS analysis is shown as a comparison. For most fields, the error of HealDA is most evident at large scales. In particular, for synoptic-scale wavenumbers around $\ell=20$, the HealDA error is larger than IFS, even for fields like Q700 and T850 where the overall RMSE had initially appeared to be similar to IFS. In contrast, at the highest wavenumbers ($\ell > 150$), HealDA exhibits smaller error than IFS for all but geopotential variables. This is the consequence of smoothing that suppresses high-wavenumber variance rather than improved representation of fine-scale structure, seen clearly from the decreased power of the HealDA analysis relative to ERA5 at these wavenumbers in Figure \ref{fig:spectra}.

In sum, this spectral analysis adds to the evidence supporting our hypothesis that $||\delta x_0||$ is too large for ML DA models, assuming ours is representative, when one considers which scales are most relevant to error growth. Furthermore, it underscores the importance of not relying solely on aggregate RMSE to understand the issues at hand.

\begin{figure}
    \centering
    \includegraphics[width=1\linewidth]{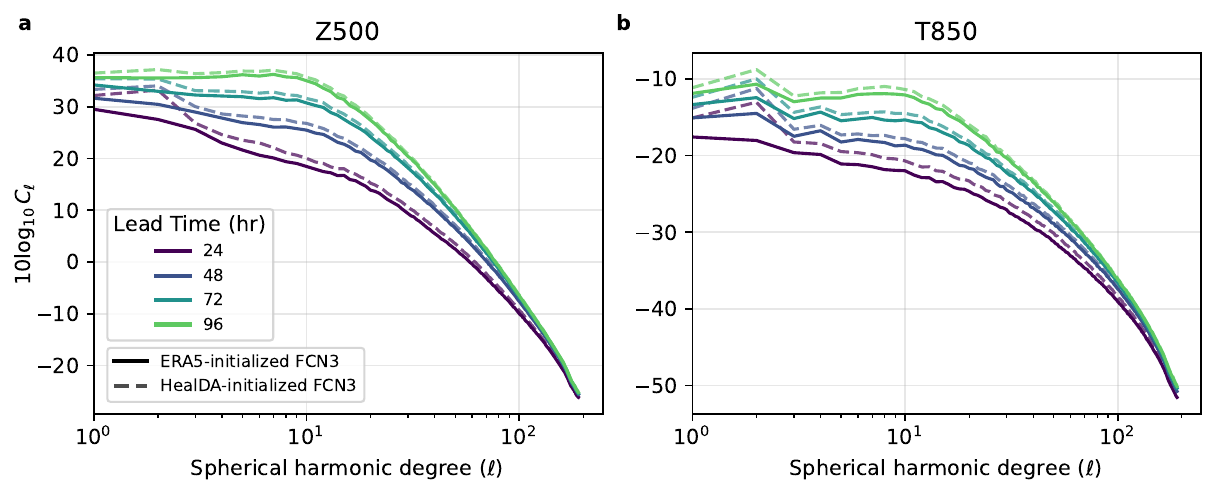}
    \caption{\textbf{Error growth.} 
   Error power spectra of FCN3 forecasts initialized with HealDA analysis versus ERA5, shown as a function of spherical harmonic degree for (a) Z500 and (b) T850 at multiple forecast lead times. Power is visualized as as $10\log_{10} C_\ell$.}
    \label{fig:error-growth}
\end{figure}

The alternative hypothesis that the error growth $\lambda+\delta \lambda $ of an ML model is modified by seeing an out-of-distribution initial condition is not strongly supported. Figure \ref{fig:error-growth} shows that in between the initial spin-up period (24 hr) and the nonlinear saturation of errors (1 week), the error is growing no faster with ML DA initialization than with ERA5, i.e. $P(\text{Our init}) \approx c P(\text{ERA5 init})$ with the same proportionality constant $c$ for both 48 and 72 hour lead times. Therefore, consistent with our observations surrounding Figure \ref{fig:fcn3-healda-vs-era5}, it seems the error is growing at roughly the same rate, but our DA simply has too much initial error at the synoptic scale.

A next important and novel observation is that \textit{ML DA displays clear overfitting at those same large-scales where its validation error is too large} to permit forecasts as skillful as conventional initialization. The difference between solid and dashed lines in Figure \ref{fig:spectra-t0} shows a striking difference between training and validation scores at large scales. This is likely caused of the global nature of the DA problem---since it is fundamentally an inverse problem that integrates information from sparse observations scattered in time and space. For an intrinsically global problem, it is easy for an over-parametrized model to simply memorize the few independent degrees of freedom present in the training data. Similar overfitting has been seen when training diffusions at large noise levels \citep{Brenowitz2025-sm}, which is similarly a global task.  20 years of data is only 7300 days, whereas deep learning models are typically trained on millions to billions of truly independent samples. 

By comparison, auto-regressive models have shown much less tendency to overfit given the locality of the task they are trained on, though they still struggle with overfitting in dynamically slow regions like the ocean \citep{Duncan2025-op} or stratosphere \citep[see SI I.2]{Watt-Meyer2024-fs}. After all, the hyperbolic nature of atmospheric flow imposes an absolute limit on how far information can travel per timestep. The fastest signals supported by the Euler equations governing the atmosphere are sound waves with a phase speed of $\SI{340}{\meter\per\second}$, while the dominant error growth is driven by synoptic scale (1000 km) processes which move more slowly $\sim \SI{50}{\meter\per\second}$. When integrated over 6 hours, this corresponds to a 1000 km radius domain of dependence. Therefore, each $(1000 km)^2$ grid box could be considered a roughly independent sample when training an auto-regressive model. This is one reason why training a DA model with a prior state or forecast as an input is a good idea. From a spectral perspective, the background prior leads to analysis increments with a flatter power spectrum than the full analysis field, which may reduce the tendency to overfit the largest scales. Aside, we did experiment with this in initial work, but found that when cycled, it only resulted in a modest 10\% improvement to our initial conditions, even after substantial tuning, which was not enough to close the gap with ERA5 initial conditions and ultimately distracting to our fundamental message here regarding an overfitting barrier thwarting skill for the state estimation task on the synoptic scale. Moreover, other works training with background still struggled with overfitting \citep{ni2025huracan}. We therefore felt that training with background was not worth the added complexity, so we removed it in our final approach reported here.

\subsection{Aurora and FengWu from HealDA}
\label{sec:deterministic-forecasts}

\begin{figure}
    \centering
    \includegraphics[width=1\linewidth]{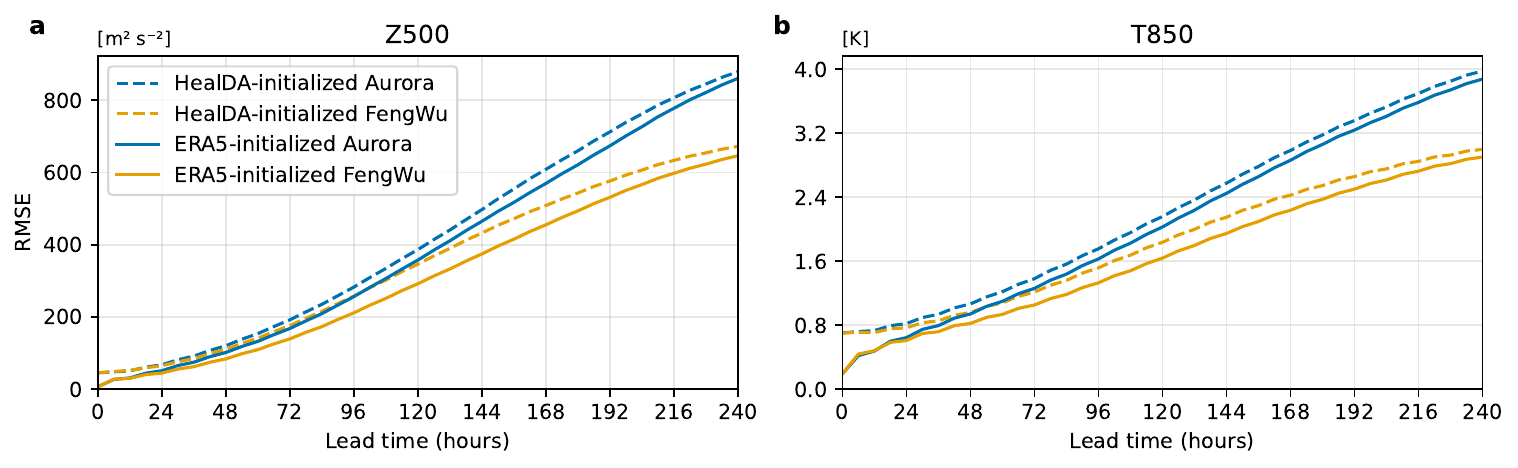}
    \caption{\textbf{HealDA can initialize FengWu and Aurora.} RMSE of deterministic Aurora and FengWu forecasts initialized from either ERA5 (solid) or HealDA (dashed). Scores are averaged over 128 initial conditions at 06/18 UTC in 2022 and verified against ERA5 on the HPX64 grid.}
    \label{fig:deterministic-comparison}
\end{figure}

Figure~\ref{fig:deterministic-comparison} expands beyond our FCN3 testbed by comparing the effect of HealDA versus ERA5 initializations in two recent state-of-the-art deterministic models that are compatible with HealDA's output variables: Aurora and FengWu. For Aurora, we use the pretrained $0.25^\circ$ checkpoint. For FengWu, we use the public 13-vertical-level deterministic model.

Overall the results are consistent: ERA5- and HealDA-initialized RMSE curves for both Aurora and FengWu evolve nearly in parallel. Relative to ERA5 initialization, HealDA initializations incur only $\sim$12\,h of effective lead-time loss for Aurora and $\sim$18-24\,h for FengWu. Importantly, neither forecast model was trained on HealDA-like initial conditions, reinforcing that once analysis error is sufficiently small, the data assimilation and forecast components can be cleanly decoupled and combined in a plug-and-play manner.

\subsection{Robustness to Observing System Changes}
\label{sec:robustness}

Figure \ref{fig:analysis-rmse}B shows that our selected 2022 test period includes both sporadic gaps and extended outages for microwave sounders in the UFS Replay dataset. We surmise the sporadic dropouts may be natural, but the extended month-long dropouts may be data processing errors in the dataset publishing. Note that AMSUB is not shown because, although it was present for parts of the training period, it has been discontinued and replaced by the newer ATMS sensor in recent years. Cross-referencing against the analysis RMSE in Figure \ref{fig:analysis-rmse}A, we find that some of the large transient RMSE spikes in Z500 and particularly Q700 (Figure \ref{fig:analysis-rmse}) are very well aligned with the extended MHS/ATMS outages, such as during May 2022.

Importantly, the RMSE for most fields is relatively stable even during these dropout periods, increasing on the order of 5-10 \%\ for most fields, but causing larger variability in geopotential fields (Z500). This suggests that our approach of explicitly representing unobserved pixels with zeros and an observability mask makes HealDA reasonably robust to both sporadic and sustained sensor loss -- an attractive property for an ML DA module.

We note that in a traditional NWP DA system, the background forecast carries information forward from earlier observations and would be expected to mitigate such intermittency. By contrast, HealDA relies entirely on the current observation window, so extended sensor outages have a heightened effect on the analysis. Incorporating an explicit background in the ML approach may help reduce this sensitivity, as demonstrated in \citep{Sun2025Fuxi}.

\section{Conclusions}

In this work, we have introduced HealDA, a flexible ML architecture for initializing global weather forecasts in a direct observation-to-state framework. HealDA is computationally lightweight, producing analyses orders of magnitude faster than traditional NWP DA pipelines. The model combines a flexible encoder that treats observational data as raw point cloud data. This encoder mirrors the structure of the observing system: scalar observables are grouped into sensors, each of which is observed from multiple platforms. The output of the encoder is latent vectors defined on the HEALPix global grid, which is highly convenient for processing observations due to its hierarchical and equal-area structure. Care is taken to preserve the memory usage throughout all layers to avoid imposing any information bottlenecks. These latent vectors are then processed by a standard transformer processor and then decoded with patch-decoding. Overall, we present this architecture as a simple alternative to other point-cloud encoding schemes (e.g., SetConv), and it is surprisingly robust to changes in the observations processed by the model. We therefore expect that our encoder would likely work zero-shot, and more certainly with fine-tuning, with new sets of observations of a similar type (microwave or conventional). For example, microwave sounders with channels overlapping those seen during training, or conventional observations measuring the same physical quantities, could be ingested without architectural changes, while sensors with new characteristics could be incorporated by adding a lightweight sensor-specific embedder and fine-tuning.

Unlike prior work, we have highlighted that ML-based data assimilation methods can be developed independently from the forecast models. Partly out of convenience, but also to control various confounding factors, we have decided to use off-the-shelf forecasts such as FourCastNet3 and Microsoft's Aurora for our end-to-end forecasts.
Importantly, we compare the performance of all forecasts both when initialized from ERA5 as well as from our own initializations, using probabilistic metrics when available. Compared to initializing from ERA5, our initializations degrade accurate forecast lead times by about 12-24 hours, depending on the field. This is consistent with our finding that our initial error for the state estimation task versus ERA5 is at least 50\% larger than the ERA5-HRES difference, which we argue should serve as an upper bound for an ERA5 emulator. Nonetheless, a review of the literature reveals that our model is at the forefront of ML DA approaches both in terms of initial and forecast error, especially if one focuses on probabilistic metrics of already-smooth fields like Z500, which are harder to ``cheat'' by making the forecast or reference blurry (see Section \ref{sec:scoring-choice}). Quantitatively, using HealDA (L1), the checkpoint fine-tuned using only Level~1 inputs (see Section~\ref{sec:lvl2-ablation}), Z500 initial-condition RMSE is reduced by 20\% relative to Aardvark and 38\% relative to FuXi Weather, with corresponding reductions of 26\% and 15\% in 5-day forecast RMSE when coupled with FCN3 (Table~\ref{tab:da_comparison}).

Despite these gains, it seems clear that ML DA models still have more to learn before competing with physics-based approaches. To further improve results, our spectral analysis reveals a critical issue with overfitting, especially at large-scales and for upper-tropospheric fields. While stronger priors—such as using a background or scale-dependent regularization—can reduce this overfitting, it is clear that the ML DA task is more prone to overfitting than typical auto-regressive training.
In our opinion, the best path to improving ML DA systems is to obtain more training data. For example, ACE \citep{wattmeyer2023acefastskillfullearned} reports clear overfitting when trained on 10 years of 6-hourly simulations, which is substantially reduced when scaling to 100 years of data, suggesting data beyond $\sim$20 years of ERA5 reanalysis may be required to constrain large-scale DA errors.
Moreover, as HealDA is trained to reproduce a reanalysis, its performance is largely bound by the quality of that reanalysis. Thus, actually \emph{outperforming} physics-based analysis, if possible, will require some added information, for example, by using an inverse modeling approach and accurate observation simulator, but that is further down the road.

Another limitation is that we have left the numeric value of the true error $||\delta x_0||$ somewhat hazy and in places rely on the ERA5-HRES discrepancy as a rough proxy. It is unlikely our core conclusions about the overfitting or skill versus other ERA5 emulators will be altered by the choice of verification dataset, but future work should more precisely quantify the true error versus observations \citep[Fig. 14]{Hersbach2020ERA5}.

\section{Methods}
\label{sec:methods}
\subsection{HEALPix Grid}
\label{subsec:healpix}

We represent all global fields on the equal-area HEALPix grid \citep{gorski2005hpx}. HEALPix partitions the sphere into 12 base faces, each subdivided into an $N_{\text{side}} \times N_{\text{side}}$ square grid of pixels, where $N_{\text{side}}= 2^l$ corresponds to the resolution level $l$ of the grid. In this work we use HPX64 ($N_{\text{side}} = 64$, $l=6$), which corresponds to $\sim 1^\circ$ ($\approx 100$\,km) resolution and $N_{pix} = 12 \times 64^2 = 49{,}152$ pixels globally. The equal-area property means that simple averages over grid cells are equivalent to area-weighted (latitude-weighted) global means on a regular lat–lon grid, so no special pixel-wise weighting is necessary during training, and likewise, our global RMSE, ACC, and CRPS scores are computed as unweighted averages over HEALPix cells. The hierarchical indexing structure of HEALPix also makes it easy to coarsen or upsample and thus switch between resolutions.

Whenever we apply convolutions on the HEALPix grid, we follow the face-wise scheme outlined by \citep{karlbauer2024advancingparsimoniousdeeplearning, Brenowitz2025-sm}. Each global field is split into 12 faces, halo cells are exchanged between neighboring faces according to the HEALPix connectivity to generate padding, and then a standard 2D convolution is applied to each padded face independently. The faces are finally unfolded to recover the flattened global structure.

\subsection{System Architecture}

\begin{figure}
    \centering
    \includegraphics[width=1\linewidth]{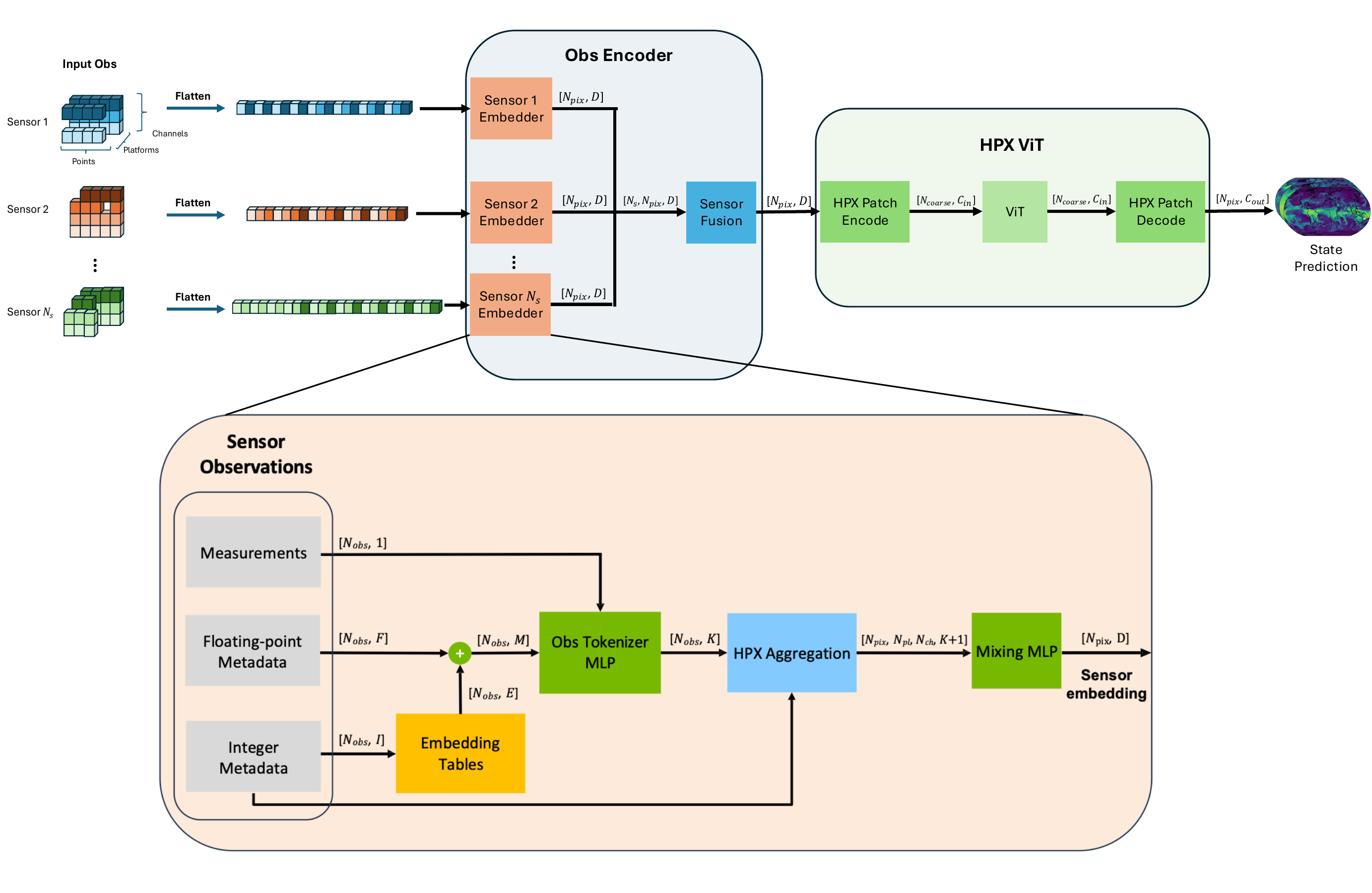}
    \caption{\textbf{HealDA network architecture}. Observation streams are flattened and then routed through sensor-specific embedders to produce per-sensor HPX feature maps of embedding dimension $D$. Each raw observation comprises integer metadata (e.g., HPX pixel, channel, platform), floating-point metadata (e.g., satellite scan angles, local solar time, pressure, height), and the measurement itself. Integer metadata are mapped through embedding tables and combined with featurized floating-point metadata along with the measurement through an Obs tokenizer MLP, yielding per-observation tokens of dimension $K$. These tokens are then scatter-reduced onto the HPX grid to form grid blocks indexed by platform and channel, and a mixing MLP produces the latent representation. All sensor maps are then fused to provide a final obs embedding, which is passed to the HPX ViT backbone. The HPX ViT consists of a 2x2 downsampling convolution, the actual ViT, and ends with a 2x2 upsampling convolution, producing the desired atmospheric fields.}
    \label{fig:healda-arch}
\end{figure}

Our system (Figure \ref{fig:system-diagram}) couples an observation-to-state DA model with an off-the-shelf forecast model. First, the DA model (Figure~\ref{fig:healda-arch}) ingests observations from a 24-hour window around the analysis time, along with auxiliary inputs (static fields such as orography and land–sea mask, and calendar features), and produces an analysis state on the HPX64 grid. The observations are processed by our observation encoder module, which maps sparse, point-cloud-like data to a dense HPX64 feature field. Calendar information (time of day, day of year) is featurized using Fourier features. All inputs are then concatenated along the channel dimension and passed through an HPX ViT, which outputs an analysis state associated with the observation window. This analysis state is used as the initial condition for off-the-shelf forecast models to generate 10-day medium-range forecasts. The DA module is trained to predict ERA5 reanalysis.

\subsection{Observation Encoder}
\label{sec:obs-encoder}
The observation encoder (Figures~\ref{fig:healda-arch}) transforms heterogeneous point-cloud-like observations from multiple sources (satellite and in-situ)  into dense features on the HPX64 grid. At the HPX64 resolution, we have $N_{\text{pix}}=49{,}152$ pixels, and for each pixel, the encoder produces a $D$-dimensional feature vector. For each sensor $s$, the encoder maps a variable-length set of raw observations to a per-sensor latent feature map $F^{(s)} \in \mathbb{R}^{N_{\text{pix}} \times D}$. Our sensor set includes microwave sounders (AMSU-A, AMSU-B, ATMS, and MHS) as well as a ``conv'' sensor that groups all conventional observations, including those from surface stations, aircraft, radiosondes, buoys, GNSS-RO, and satellite-derived wind retrievals. These per-sensor maps are then fused across sensors (Section~\ref{sec:sensor-fusion})
and passed to the HPX ViT backbone.

We operate on a flattened observation representation, where each observation corresponds to a single scalar measurement. For example, a microwave sounder that produces a 5-channel sounding yields 5 such observations, one per channel. In particular, each observation $i$ is described by
(i) a scalar measurement $y_i$;
(ii) a vector of continuous metadata containing its location and time $(lat_i, lon_i, t_i)$, and fields such as height $h_i$, pressure $p_i$, and viewing geometry (scan angle, satellite zenith, solar zenith); and
(iii) a small set of integer tags.
The integer tags include an HPX pixel index $x_i$, a channel index $c_i$, a platform (satellite) index $pl_i$, and an \emph{observation type} identifier $\tau_i$ that encodes the source (radiosonde, buoy, etc.) for conventional sources and is zero otherwise. We refer to the integer tags collectively as the integer metadata.

Given this representation, the Obs encoder proceeds in four stages:
\begin{enumerate}
    \item Remove invalid observations in a simple quality control step.
    \item Featurize the raw continuous metadata into a fixed-dimensional metadata vector of floating-point values.
    \item Tokenize each observation by combining the measurement, floating-point metadata, and integer embeddings in an Obs tokenizer MLP.
    \item Aggregate these tokens onto the HPX grid, apply a per-sensor
    mixing MLP, and fuse the resulting
    HPX feature maps across sensors.
\end{enumerate}

\subsubsection{Sensor Embedders}

\paragraph{Quality Control}
\label{sec:obs-filtering}
Before featurization, we remove invalid observations to ensure finite and physically plausible values. We drop non-finite measurements and enforce sensor and variable-specific valid ranges
(e.g., microwave radiances in $[0,400]$~K, conv specific humidity in $[0,1]$, conv wind components $(u,v)$ in $[-100,100]~\mathrm{m\,s^{-1}}$).
For conventional observations, we additionally require $\text{height}\in[0,60{,}000]$~m and
$\text{pressure}\in[200,1100]$~hPa for all but GNSS-RO, and $\text{pressure}\in[0.5,1100]$~hPa for GNSS-RO.

\paragraph{Floating-point metadata encoding}
\label{sec:float-metadata}
For each observation $i$, we construct a continuous metadata vector
$m_i \in \mathbb{R}^{D_{\text{meta}}}$ from the available scalar fields (time, location, viewing geometry, height, pressure), which is then concatenated with the integer embeddings and the measurement before the Obs tokenizer MLP. The featurization is summarized below.

\begin{itemize}
    \item \textbf{Local solar time.}
    Given longitude $\lambda_i$ (in degrees) and the absolute observation time $t_i$ (in seconds since epoch), we approximate the local solar time as 
    \[ \mathrm{LST}_i = \bigl( (t_i / 3600) + \lambda_i/15 \bigr) \bmod 24 , \]
    ignoring the equation of time. We normalize $\mathrm{LST}_i / 24$ and apply a Fourier feature mapping with two frequencies, yielding 4 features per observation ($\phi_{\text{LST}} \in \mathbb{R}^4$).

    \item \textbf{Relative time.}
    Let $t_0$ denote the target analysis time (in seconds). We compute the relative time in hours, $\Delta t_i = (t_i - t_0)/3600$, normalize it by 24 hours, and use a simple two-dimensional encoding
    \[
        \phi_{\text{time}}(\Delta t_i)
        = \bigl[\Delta t_i/24,\ (\Delta t_i/24)^2\bigr] \in \mathbb{R}^2.
    \]

    \item \textbf{Height and pressure.}
    When height $h_i$ (m) and pressure $p_i$ (hPa) are available, we map each
    quantity to $[0,1]$ by linear rescaling with clipping,
    \[
        \tilde{h}_i = \mathrm{clip}\!\left(\frac{h_i}{h_{\max}},\, 0,\, 1\right),
        \qquad
        \tilde{p}_i = \mathrm{clip}\!\left(\frac{p_i}{p_{\max}},\, 0,\, 1\right),
    \]
    using $h_{\max}=60000\ \mathrm{m}$ and $p_{\max}=1100\ \mathrm{hPa}$.
    We then apply Fourier features with four frequencies to each normalized value,
    $\phi_{\text{ht}}(\tilde{h}_i) \in \mathbb{R}^8$ and
    $\phi_{\text{prs}}(\tilde{p}_i) \in \mathbb{R}^8$.
    If height or pressure are not provided for a given observation (e.g., for satellite-only data), we insert NaNs for the corresponding slots.

    \item \textbf{Scan angle and viewing geometry.}
    For microwave sounders, we encode the scan angle
    $\xi_i$ via a polynomial feature
    \[
        \phi_{\text{scan}}(\xi_i)
        = \bigl[\xi_i/50,\ (\xi_i/50)^2\bigr] \in \mathbb{R}^2,
    \]
    which roughly maps the typical scan-angle range to $[-1,1]$. The satellite zenith angle $\theta^{\text{sat}}_i$ is encoded through $\cos$ features,
    \[
        \phi_{\text{sat}}(\theta^{\text{sat}}_i)
        = \bigl[\cos\theta^{\text{sat}}_i,\ \cos^2\theta^{\text{sat}}_i\bigr]
        \in \mathbb{R}^2,
    \]
    and the solar zenith angle $\theta^{\odot}_i$ through
    \[
        \phi_{\text{sun}}(\theta^{\odot}_i)
        = \bigl[\cos\theta^{\odot}_i,\ \sin\theta^{\odot}_i\bigr]
        \in \mathbb{R}^2.
    \]
    For conventional observations without viewing geometry information, we fill these entries with NaNs.
\end{itemize}

We concatenate all of the above features to obtain $m_i \in \mathbb{R}^{D_{\text{meta}}}$ (28 dimensions in our configuration), and finally replace all NaNs with zeros. This continuous metadata vector is passed to the Obs tokenizer MLP as input.

\paragraph{Obs Tokenizer MLP}
\label{sec:obs-tokenizer}
Each observation $i$ is represented by a scalar measurement $y_i$, a floating-point
metadata vector $m_i \in \mathbb{R}^{D_{\text{meta}}}$, and integer metadata, including an observation-type identifier $\tau_i$. The Obs tokenizer maps these inputs to a $K$-dimensional token $z_i \in \mathbb{R}^K$.

We embed the observation type $\tau_i$ via a learnable embedding table, $e^{\text{type}}_i \in \mathbb{R}^{D_{\text{emb}}}$, and form a concatenated input vector
\[
x_i = \bigl[\, y_i,\ m_i,\ e^{\text{type}}_i \,\bigr]
    \;\in\; \mathbb{R}^{1 + D_{\text{meta}} + D_{\text{emb}}}.
\]
This vector is passed through a 2-layer MLP with LayerNorm and SiLU activation, and the MLP output is concatenated with the raw measurement to get the final token as
\[
z_i = \bigl[\, y_i,\ \mathrm{MLP}(x_i) \,\bigr]
\in \mathbb{R}^K,
\]

We embed only the observation type. Platform– and channel–specific structure is already available to the model at the grid-aggregation stage: the mixing MLP receives, for each pixel, a full block of features indexed by platform and channel, and can therefore learn arbitrary platform/channel-specific transformations without requiring separate embeddings in the tokenizer.

\paragraph{HPX Aggregation and Mixing MLP}
\label{sec:sensor-aggregation}

Let $N_{\text{pix}}$ denote the number of HPX pixels at the aggregation resolution. For a given sensor $s$, let $N_{\text{pl}}$ be the number of platforms carrying that sensor (e.g.\ MetOp-A/B/C  for microwave sounders; for the conv sensor, this is set to $1$) and $N_{\text{ch}}$ the number of channels. For microwave instruments, channels correspond to the standard channels (frequency bands) of each instrument (e.g.\ 15 AMSU-A channels,
5 MHS channels). For the conv sensor, channels index 1 of 8 different conventional streams: surface pressure,  specific humidity, temperature, GNSS-RO bending angle and corresponding retrieved $t$/$q$ profiles (3 channels), and $u/v$ winds (2 channels).

Let $z_i \in \mathbb{R}^K$ denote the token for observation $i$, with associated HPX pixel $x_i$, platform $pl_i$, and channel $c_i$. We aggregate tokens by computing, for each pixel $x$, platform $pl$ , and channel $c$,
\[ H_{x,pl,c} = \operatorname{Mean}\{ z_i \mid x_i = x,\ pl_i = pl,\ c_i = c \}, \] and empty bins are set to zero. This can be implemented in PyTorch using a scatter--reduce operation and yields a tensor $H \in \mathbb{R}^{N_{\text{pix}} \times N_{\text{pl}} \times N_{\text{ch}} \times K}$, i.e.\ a $K$-dimensional feature vector at each HPX pixel $x$ for every (platform, channel) pair of that sensor. Although aggregation uses a simple mean, temporal information is preserved through time-conditioned observation embeddings, so the reduction operates on time-aware latent features rather than raw measurements. More expressive temporal aggregation schemes may, nevertheless, further improve performance.

Alongside $H$, we construct a per-(pixel, platform, channel) observability mask $M \in \{0,1\}^{N_{\text{pix}} \times N_{\text{pl}} \times N_{\text{ch}}}$, with $M_{x,pl,c} = 1$ if there exists an observation $i$ such that $x_i = x,\ pl_i = pl,\ c_i = c$, and $M_{x,pl,c}=0$ otherwise. This allows the network to distinguish between observed and unobserved pixels. We stack this mask onto the features, and then flatten the platform, channel, and feature dimensions so that $[H, M]$ becomes a single vector in $\mathbb{R}^{N_{\text{pl}} N_{\text{ch}} (K+1)}$. A mixing MLP acts on this per-pixel vector to produce $f_x \in \mathbb{R}^D$, a $D$-dimensional latent feature at pixel $x$. Across all pixels, this yields a per-sensor HPX feature map \[F \in \mathbb{R}^{N_{\text{pix}} \times D}, \] which is then passed to the downstream sensor fusion module.

\subsubsection{Sensor fusion}
\label{sec:sensor-fusion}

For each sensor $s \in \{1,\dots,S\}$, the corresponding sensor embedder produces a latent feature map $F^{(s)} \in \mathbb{R}^{N_{\text{pix}} \times D}$ on the HPX grid,  where $N_{\text{pix}}$ is the number of HPX pixels and $D$ the feature dimension. We fuse these into a single HPX representation via a simple uniform reduction,
\[
    F = \frac{1}{\sqrt{S}} \sum_{s=1}^S F^{(s)},
\]
which preserves the scale of the features as the number of sensors varies. In preliminary experiments, this uniform fusion performed as well as more complex schemes, so we adopt this simpler method. The fused representation $F$ is then passed to the HPX ViT.

\subsection{HPX ViT}

We use a 330M-parameter ViT-style vision transformer adapted from \citep{von2022diffusers} as our backbone, augmented with HEALPix-specific $2\times 2$ patch-encode and patch-decode layers along with positional embeddings. In particular, the patch-encode is a $2\times 2$ convolution, while the patch-decode is a $2\times 2$ transposed convolution. The architecture uses 24 transformer blocks and an embedding dimension of 1024, based on the DiT-L settings \cite{peebles2023scalablediffusionmodelstransformers}. Although the backbone follows a DiT-style architecture, HealDA is used purely as a deterministic regression model rather than a diffusion model. We retain the DiT conditioning structure (adaptive normalization), which produces per-block scale, shift, and gating mechanisms. However, no stochastic noise or class conditioning is provided during training or inference; the corresponding conditioning inputs are fixed to zero (see \ref{sec:vit-conditioning} for more details on the conditioning layers). Within the attention module, we apply RMS normalization to the query and key vectors before the dot-product \citep{zhang2019rmsnorm}, with element-wise affine parameters disabled. This normalization is used for numerical stability when training in bfloat16 precision; without this, we found the later stages of training to be unstable.

\subsection{Forecasting pipeline}
We regrid our HPX64 initial conditions (both HealDA and ERA5) to the 0.25-degree resolution of the forecasting models (see Section \ref{sec:scoring-procedure}). We perform our evaluations on the resultant forecast down-sampled to the HPX64 resolution. 

\subsection{Training}

HealDA is a deterministic model trained to predict ERA5 reanalysis targets from raw observational input. We use 2000--2021 as the training period, as the observational record for microwave sounders is sparse prior to 2000, and reserve 2022 for testing. The observation encoder and HPX ViT backbone are trained together using a Huber regression loss on ERA5 targets. We use the AdamW optimizer \citep{loshchilov2019decoupled}, with $\beta_1=0.9$, $\beta_2=0.95$, and a weight decay of 0.05.
We use a base learning rate of $5\times10^{-4}$ for the HPX ViT backbone and $1\times10^{-4}$ for the observation encoder, with a linear warm-up over the first 50k samples followed by cosine decay to zero over 10M samples, corresponding to approximately 333 epochs in total. Additional regularization is applied via dropout and stochastic depth (drop path). We use a dropout probability of $p=0.05$. Following \citet{huang2016stochasticdepth}, we employ a linearly increasing drop-path schedule across transformer blocks, with drop probability increasing from 0 in the initial transformer block to a maximum value of 0.1 in the final layer. Training is performed on a single H100 node (8 GPUs) with a total batch size of 8 in bfloat16 precision. Table~\ref{tab:hyperparams} summarizes these settings.

\subsection{Compute and Memory Requirements}
\paragraph{Training}
HealDA is trained on a single node with 8 NVIDIA H100 (80\,GB) GPUs,
requiring approximately 1,600 GPU-hours in total (8.3 days wall-clock) and the full 80GB memory of each GPU.

\paragraph{Inference}
HealDA can be inferenced on a single H100 GPU to produce a global analysis in 70ms. CPU-side data loading and preprocessing, however, takes $\sim 0.7\,\text{s}$ per analysis, so the full inference pipeline can be run in under one second. Inference requires only $\sim$20\,GB of GPU memory, allowing deployment on GPUs beyond H100-class hardware.

\subsection{Data}
\label{sec:data}

\subsubsection{State estimates}

\paragraph{ERA5} As an ML target, we employ state estimates from ECMWF Reanalysis version 5 (ERA5). ERA5 is computed by filtering observations from the historical record using a fixed version of ECMWF's cycled data assimilation scheme \citep{Hersbach2020ERA5}.
This process unrolls in 12-hour data assimilation (DA) windows from 09 -- 21 UTC and 21 -- 09 UTC. Within each window, the 4DVar procedure solves for an initial condition at the beginning of the window, which minimizes a cost function combining the mismatch of observations and the background error of a background forecast. Therefore, all state estimates within each DA window see observation up to the end of the window. This means that a 09 UTC ERA5 state estimate sees observations up to 12 hours ahead, which is an important effect to consider when using DA as initial conditions.

ERA5 climatology is downloaded from  WeatherBench2 \citep{rasp2024weatherbench2}.

\subsubsection{Observational Datasets}
\label{sec:ufs-replay}

We obtain observation data from the UFS Replay archive \citep{UFSReplay}. In addition to analysis states (which we don't use for this work), this dataset contains many of the observations used by the NOAA operational forecast systems and broadly matches the observing systems used to produce modern reanalyses (e.g., ERA5), although the observations in the UFS Replay are thinned to $1^\circ$ spatial resolution. It has proved a convenient and comprehensive archive of observation data in netCDF format. We use the non-bias corrected observations stored in the \texttt{Observation} field. We used the following types of observational data. The impact of removing Level~2 products, including satellite wind retrievals and GNSS-RO temperature and humidity channels, is analyzed in Section~\ref{sec:lvl2-ablation}. \citet{Healy} surveys how these observation streams impact forecast accuracy. 

\paragraph{Microwave Sounders} Microwave sounders are satellite-borne instruments that measure upwelling microwave radiances at frequencies sensitive to the vertical distribution of temperature and humidity in the atmosphere. These frequencies are selected to sample absorption by atmospheric constituents, primarily $H_2O$ and $O_2$, resulting in coarse vertical weighting functions. To optimize for broad spatial coverage at a reasonable temporal frequency, these sensors are deployed in polar orbits. Microwave radiances are generally robust in cloudy conditions and have relatively low data volume compared to hyperspectral infrared observations. For these reasons, and because microwave sounders already provide good global coverage, we restrict our satellite radiance inputs to microwave sensors in this work. Specifically, we use observations from AMSU-A, AMSU-B, ATMS, and MHS aboard the NOAA-15–20, Metop-A–C, and Suomi-NPP platforms.

\paragraph{Conventional observations} The archive includes direct \emph{in situ} observations of humidity, temperature, pressure, and winds from a variety of sources, including aircraft, radiosonde, and surface observations. Given the expense of such data collection, its spatial coverage is sparse and concentrated over land and developed countries.

\paragraph{Satellite Winds} The satellite winds include scatterometer (e.g., ASCAT) and atmospheric motion vector (AMV) data. Scatterometer winds are inferred from microwave radar backscatter measured over the ocean surface, which depends on wind-driven surface roughness and is related to the 10-m ocean wind through a geophysical model function (GMF). Measurements over land are excluded using a land–sea mask, and the GMF inversion can yield multiple wind vector solutions within each wind vector cell. An ambiguity-removal step selects the solution most consistent with spatial constraints and a first-guess NWP field interpolated to the observation location and time \citep{ascat-manual,ecmwf-seminar}. 

AMVs are produced by tracking clouds or coherent water-vapor features through successive satellite imagery and interpreting their displacement as an estimate of the atmospheric wind vector \citep{Forsythe2008}. In addition to feature-tracking uncertainty, much of the error stems from the background-dependent height assignment, where tracked features are assigned a representative height using observed radiances, radiative-transfer models, and short-range forecast temperature and moisture profiles. Nevertheless, AMVs provide important wind information, especially in regions where other wind observations are sparse. While both scatterometer and AMV winds are Level~2 products, we include them as they can be available operationally.

\paragraph{GNSS-RO} Global Navigation System Radio Occultation is a remote sensing method that measures how a radio signal's path is bent as one satellite orbits away from a direct line of sight as the Earth's limb appears between them \citep{ecmwf_gpsro_2015}. This is called an occultation event, and as the radio signal between the two satellites passes through different layers of the atmosphere, the local humidity and temperature impact the index of refraction. We use the raw Level~1 bending angle product as well as temperature and humidity channels provided in the dataset. As discussed in Section~\ref{sec:lvl2-ablation}, these temperature and humidity channels are Level~2 products derived within the GSI diagnostic pipeline and are not direct GNSS-RO measurements.

\subsection{Forecast baselines}

IFS ENS and HRES forecast data at the 0.25° resolution are downloaded from WeatherBench2 \citep{rasp2024weatherbench2}. Forecast model inferencing of FCN3, Aurora, FengWu, and Pangu-Weather is performed with the NVIDIA Earth-2 Studio software package \citep{earth2studio_v0.9.0}. We interpolate the initial conditions for all models from HPX64 to the 0.25 degree grid to keep the comparison clean, and also out of convenience.

\subsection*{Code Availability}

Code and model weights are available through the \href{https://github.com/NVIDIA/physicsnemo}{PhysicsNemo} and \href{https://github.com/NVIDIA/earth2studio}{Earth2Studio} packages.

\subsection*{Acknowledgements}

We thank Jaideep Pathak, Suman Ravuri, and Tung Nguyen for helpful discussions. This work has greatly benefited from the work of the Atlas and FourcastNet teams at NVIDIA, who we thank for creating fast, accurate, and well-validated forecast models. We thank the Earth2Studio \citep{earth2studio_v0.9.0} authors and, in particular, Nicholas Geneva for creating a flexible and easy to use inference package with support for various models. These open NVIDIA products have made our paper immeasurably stronger. We would also like to acknowledge the helpful comments left by anonymous NOAA and MITRE reviewers. A portion of this work was supported by the MITRE Independent Research and Development Program.

\subsection*{Supplementary Materials}
\textbf{This PDF file includes:}

Supplementary text

Figs. SI1 to SI15

Tables SI1 to SI2

\bibliography{references}
\bibliographystyle{unsrtnat}

\appendix
\clearpage

\section{Supplementary Information}
\renewcommand{\thefigure}{SI\arabic{figure}}
\setcounter{figure}{0}
\renewcommand{\theHfigure}{SI\arabic{figure}}

\renewcommand{\thetable}{SI\arabic{table}}
\setcounter{table}{0}
\renewcommand{\theHtable}{SI\arabic{table}}

\subsection{ViT Conditioning Layers}
\label{sec:vit-conditioning}

Despite training a deterministic model, our implementation adopted the adaptive modulation design from the DiT architecture, specifically the AdaLayerNormZero (AdaLN-Zero) formulation, in which a conditioning MLP produces per-block scale, shift, and gating parameters. We used a zero scalar for the noise and an empty tensor (dimension 0) for the conditioning. Because we retained the default DiT embedding configuration, the conditioning embedding width was set to $4\times$ the hidden dimension, and the noise channels was set to the hidden dimension. This yielded
\texttt{noise\_channels=1024}, \texttt{condition\_dim=0}, and \texttt{emb\_channels=4096} for our hidden dimension of $d=1024$.

Each transformer block, therefore, included a modulation MLP mapping $4096 \rightarrow 6d$, resulting in approximately $4096 \times 6144 \approx 25$M parameters per block. Across 24 blocks, this contributed $\sim 600$M parameters solely for adaptive normalization, increasing the total parameter count to approximately 1B. Empirically, removing adaptive modulation entirely slowed training convergence, indicating the additional parameters proved useful for optimization dynamics. However, reducing the conditioning embedding width (e.g., using \texttt{noise\_channels = emb\_channels = 128}) preserved comparable convergence (aside from an initial spin-up phase), reducing the modulation parameter overhead to only a few million parameters.

Because both the noise input (scalar zero) and conditioning input (empty tensor) are constant, the outputs of the modulation MLPs are constant across all samples. To reduce the number of parameters, after training, we evaluate each modulation MLP once at the fixed conditioning input and replace the learned linear mapping $Wc + b$ with the resulting constant vector $Wc_0 + b$. This eliminates the modulation weight matrices and associated matrix multiplications. The resulting post-training reparameterization reduces the checkpoint size from $\sim$1B to 330M parameters without altering the network's learned input-output mapping.

Finally, while AdaLN-Zero typically employs zero initialization to ensure the modulation initially acts as an identity mapping, we instead used PyTorch's default linear initialization for the modulation MLPs.

\subsection{Scoring Procedure}
\label{sec:scoring-procedure}

We perform all verification on the HPX64 grid. Forecast fields produced on the native $0.25^\circ$ lat--lon grid are first bilinearly interpolated onto HPX256 (HEALPix level~8), and then coarsened to HPX64 by block-averaging each $4 \times 4$ patch of HPX256 cells. The same regridding procedure is used to construct our ERA5 HPX64 reference dataset. This two-step ``interpolate then average'' procedure is more conservative than a single bilinear interpolation directly to HPX64, and helps to prevent aliasing of higher frequencies onto the HPX64 resolved modes. 

As noted in Section~\ref{subsec:healpix}, the HPX64 grid is equal-area, so global averages are computed as unweighted means over HPX64 cells when scoring. For ensemble forecasts, we report fair finite-ensemble probabilistic scores: CRPS is computed using a fair finite-ensemble estimator \citep{zamo2018crps}, and the RMSE of the ensemble mean is debiased using the sample ensemble variance. For RMSE and spread, we average MSE and ensemble variance over space and initial times, and take the square root after averaging as in \citep{rasp2024weatherbench2}.

\subsection{Choice of Scoring Reference}
\label{sec:scoring-choice}

\begin{figure}
    \centering
    \includegraphics[width=1.0\linewidth]{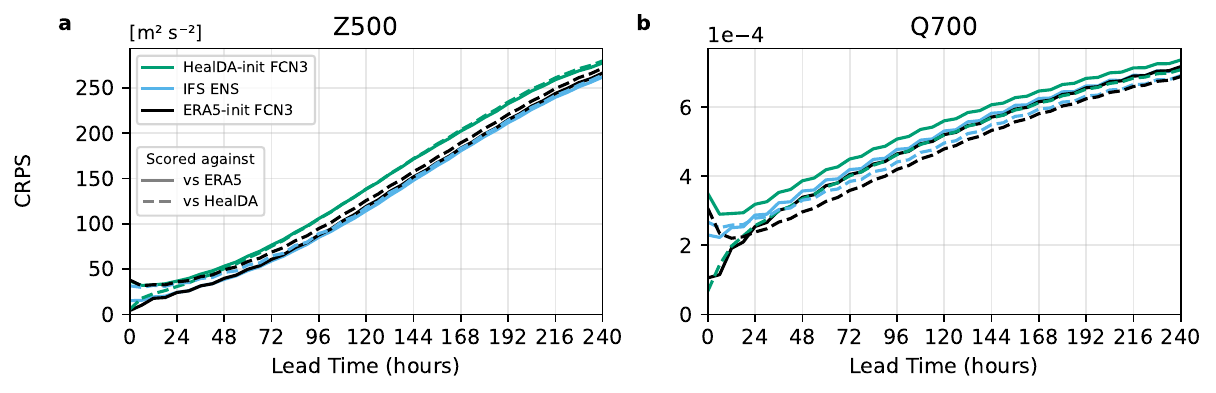}
    \caption{
    \textbf{Effect of scoring reference for different forecast systems}. 
    CRPS for multiple forecast systems (IFS ENS, ERA5-initialized FCN3, and HealDA-initialized FCN3) when verified against ERA5 (solid lines) and against the HealDA analysis (dashed lines), shown for (a) Z500 and (b) Q700. Scores are computed on the HPX64 grid for 00/12\,UTC initial times in 2022.
    }
    \label{fig:vs-analysis}
\end{figure}

In this section, we show that a common practice of scoring against one's own analysis can artificially bias the results in favor of the forecasts whose initial condition is blurrier. By definition, a score like CRPS is only proper when defined with respect to a single, common verification reference $y$ that is held fixed across all systems being compared \citep{Gneiting2007-rq}. When scoring against different verifications, we cannot conclude that a lower CRPS means a more accurate forecast. To see why, note that CRPS=0 is achieved in the degenerate case where both the ensemble $x_i$ and verification of choice $y$ are climatology ($x_i=y=\mu$). But this does not mean that climatology is a good forecast. So while the practice of verifying against one's own analysis may be the WMO standard \citep{wmo2023wmo485} for physics-based models, it is misleading to do so for ML models prone to producing blurry analyses.

Figure~\ref{fig:vs-analysis} shows how changing the verification reference alone can shift apparent ensemble skill by nearly a day. For specific humidity fields (Q700), when all systems are verified against ERA5, IFS ENS and FCN3 initialized from ERA5 have similar CRPS, and FCN3 initialized from HealDA lags by roughly 18\,h of effective lead time. When we instead verify all three ensembles against the HealDA analysis, \emph{every} system appears substantially better: the CRPS curves shift downward to the equivalent of 12-24h of effective lead time gain. This reflects the fact that HealDA’s analyses are smoother than ERA5 in these fields, with reduced small-scale variance. Scores such as CRPS and RMSE are systematically smaller when measured against this smoother target, regardless of the forecast model (FCN3 or IFS ENS) or the initial condition.

By contrast, for geopotential fields (Z500), HealDA has excess small-scale noise relative to ERA5 (see Section~\ref{sec:spectra}), and the effect largely reverses: verifying against the HealDA analysis leaves CRPS for all systems essentially unchanged or very slightly worse, rather than artificially improved. In other words, Z500—which is dominated by large scales and less sensitive to small-scale smoothing—is much harder to “cheat” via a blurrier analysis, whereas temperature and humidity fields with small-scale structure can gain an apparent 12–24\,h of lead time purely from verifying against a smoother target. 

Together, these results demonstrate that scoring a forecast against its own ML-produced analysis can substantially inflate apparent skill whenever that analysis has reduced variance at scales that remain resolved on the verification grid. Because these shifts are on the same order (12–24\,h) as the differences reported between recent ML DA systems, we argue that cross-system comparisons should use a common, fixed non-ML reference (e.g., ERA5, operational analysis, or observations), with scores against a model’s own analysis reserved for supplemental diagnostics rather than headline claims.

\subsection{ERA5- vs HealDA- initialized FCN3 Scorecards}
\label{sec:scorecards}

\begin{figure}
    \centering
    \includegraphics[width=1.0\linewidth]{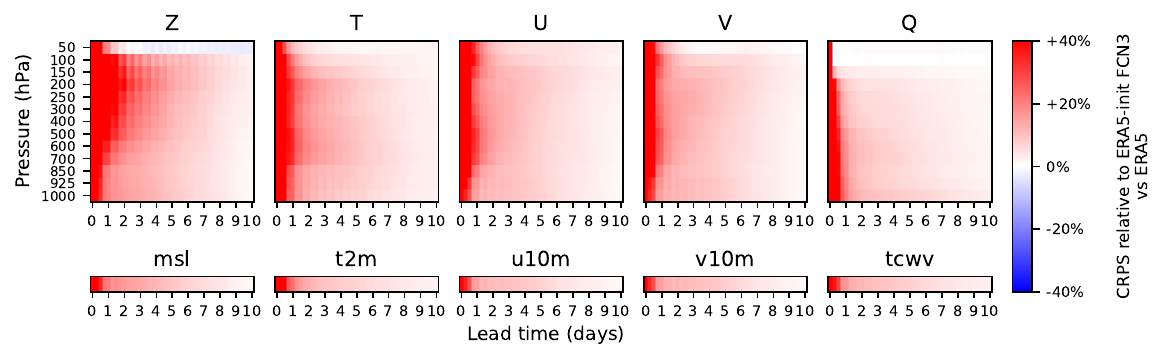}
    \caption{
    \textbf{FCN3 scorecard of HealDA vs ERA5 initialization, verified against ERA5.}
    Relative CRPS differences for FCN3 forecasts initialized from HealDA analyses versus ERA5 HPX64 analyses, with verification against ERA5 at 06/18 UTC initializations in 2022. Panels show geopotential (Z), temperature (T), winds (U, V), and specific humidity (Q) across 13 pressure levels, as well as mean sea level pressure (msl), 2\,m temperature (t2m), 10\,m winds (u10m, v10m),  and total column water vapor (tcwv). Colors indicate the CRPS of HealDA–initialized FCN3 relative to ERA5–initialized FCN3: negative values (blue) mean HealDA is better, positive (red) worse. Values are clipped to $\pm 40\%$ for visual clarity.
    }
    \label{fig:fcn3-vs-era5-scorecard}
\end{figure}

\begin{figure}
    \centering
    \includegraphics[width=1.0\linewidth]{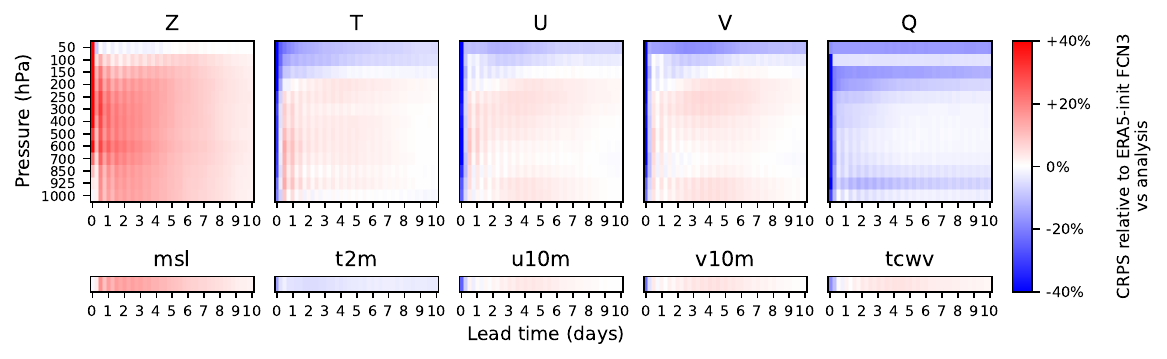}
    \caption{
    \textbf{FCN3 scorecard of HealDA vs ERA5 initialization, verified against HealDA.}
    As in Figure~\ref{fig:fcn3-vs-era5-scorecard}, but verifying each system against its analysis. HealDA–initialized FCN3 is verified against HealDA analyses, and ERA5–initialized FCN3 forecasts are verified against ERA5. Relative CRPS is again shown as a percentage and clipped to $\pm 40\%$.
    }
    \label{fig:fcn3-vs-healda-scorecard}
\end{figure}

Figures~\ref{fig:fcn3-vs-era5-scorecard} and~\ref{fig:fcn3-vs-healda-scorecard} summarize the relative CRPS of FCN3 forecasts initialized from HealDA versus ERA5 across variable and lead time combinations. When verified against ERA5 (Figure~\ref{fig:fcn3-vs-era5-scorecard}), HealDA initialization incurs an increase in CRPS relative to ERA5 for most variables and levels, with the largest losses in geopotential fields, consistent with the larger Z500 analysis errors seen in Section \ref{sec:analysis-rmse}. The penalty is most pronounced at the shortest lead times, where ERA5-initialized forecasts are being verified against their own analysis fields, while HealDA-initialized forecasts are compared against a different analysis, so any mismatch in the initial state is most visible during the first 48~h.

When we instead verify each forecast against its own analysis (Figure~\ref{fig:fcn3-vs-healda-scorecard}), the pattern changes markedly. For temperature and especially humidity, regions in the upper troposphere and lower stratosphere flip sign and show apparent \emph{improvements} (blue shading) of up to $\sim 20$\% CRPS for HealDA-initialized forecasts. This reflects the fact that HealDA analyses are smoother than ERA5 for these fields, particularly in the 50--200~hPa region, so HealDA+FCN3 forecasts look better when scored against the blurred reference, matching the discussion in Section~\ref{sec:scoring-choice}. In contrast, the Z panels change little between the two figures: geopotential is less affected by small-scale smoothing, and HealDA has some excess high-wavenumber noise, so verifying against HealDA does not systematically inflate Z skill and can slightly worsen it at longer leads. We additionally see relative skill improvement at early leads, during which the forecast would be most similar to the analysis of its initial condition.

These scorecards therefore make explicit that the choice of verification analysis can move CRPS by tens of percent---equivalent to $\sim12$--$24\text{ h}$ of apparent lead time as shown in Section \ref{sec:scoring-choice}---for many variables, while Z is comparatively robust.

\subsection{Extended Metrics}
\label{sec:extended-metrics}
\begin{figure}
    \centering
    \includegraphics[width=1.0\linewidth]{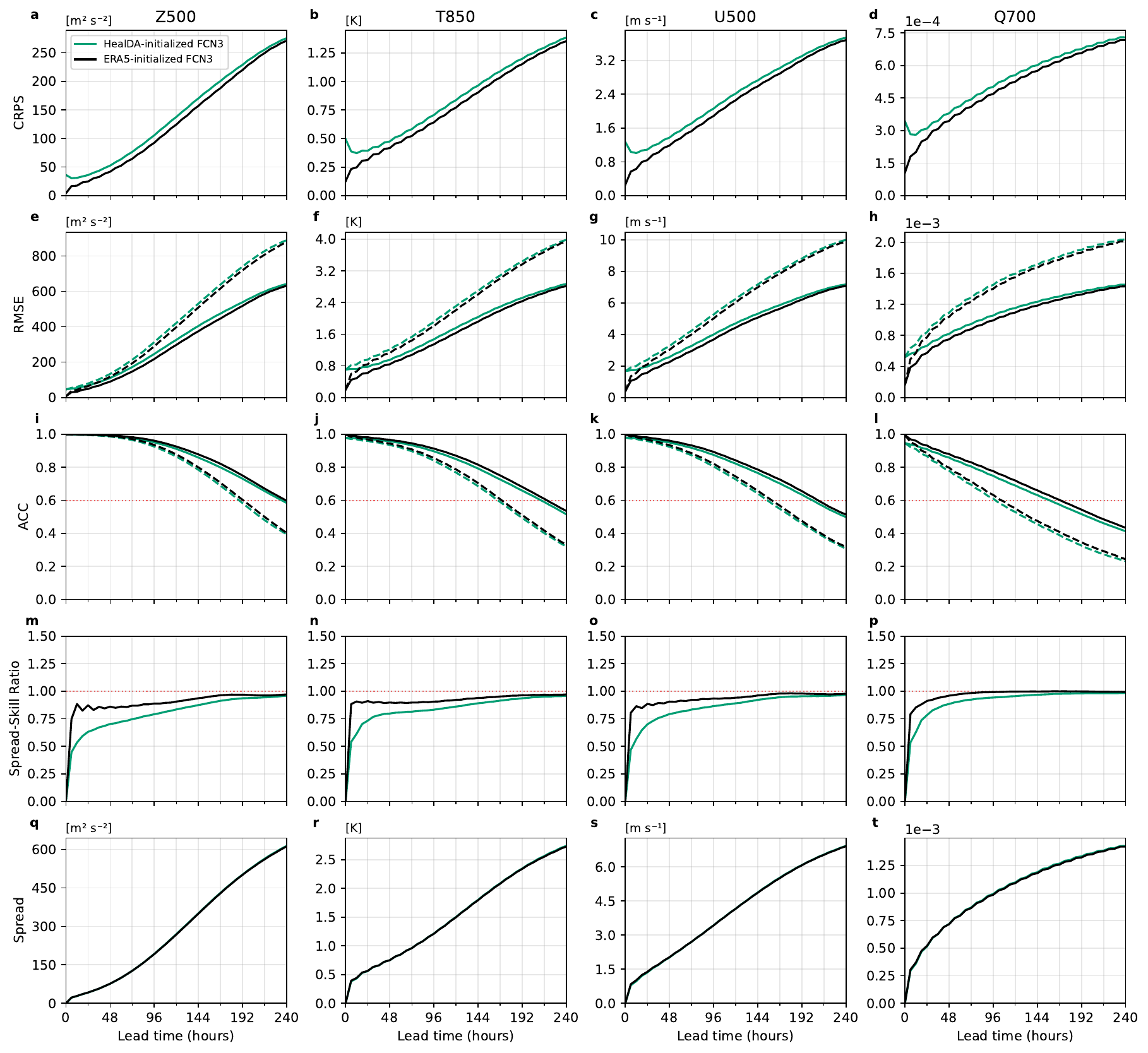}
      \caption{\textbf{HealDA- and ERA5-initialized FCN3 forecast skill.} Forecast skill of FCN3 forecasts initialized from HealDA analyses and ERA5 analyses, verified against ERA5. Solid lines indicate ensemble skill, while dashed lines indicate single member skill. Scores are averaged over forecasts initialized at 06/18 UTC in 2022. Red dotted lines mark reference thresholds: ACC exceeding 0.6 indicates skillful forecast lead time, and SSR = 1 indicates perfect spread–skill calibration.}
      \label{fig:healda-vs-era5-full}
\end{figure}

\begin{figure}
    \centering
    \includegraphics[width=1.0\linewidth]{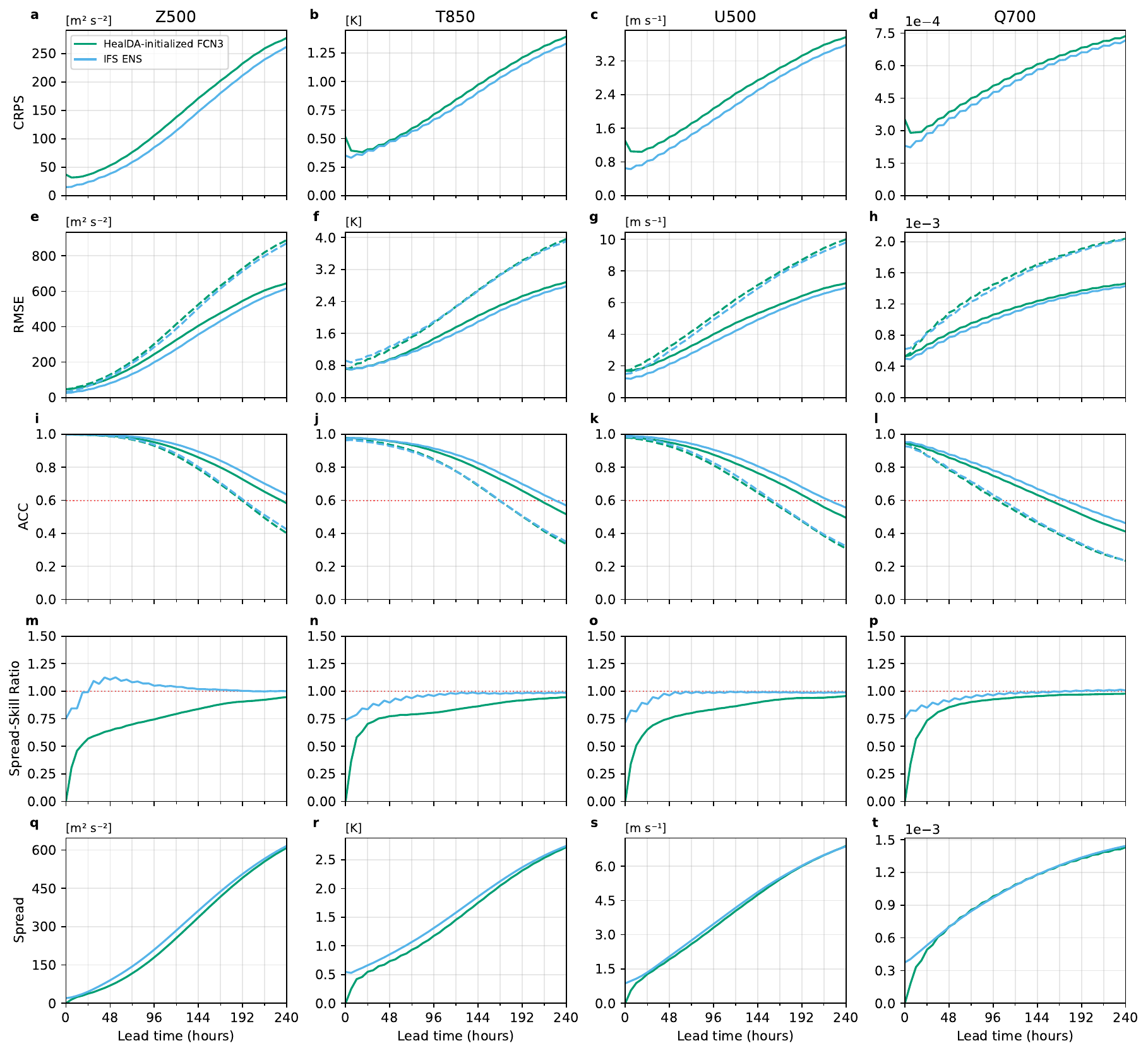}
    \caption{\textbf{HealDA-initialized FCN3 vs IFS ENS forecast skill.}
Forecast skill of HealDA-initialized FCN3 forecasts and ECMWF IFS ensemble forecasts verified against ERA5. Solid lines indicate ensemble skill, while dashed lines indicate single member skill. Scores are averaged over forecasts initialized at 00/12~UTC across 2022. Red dotted lines mark reference thresholds (ACC = 0.6; SSR = 1).}
    \label{fig:healda-vs-ifs-full}
\end{figure}

In the main text, we focus on CRPS to compare probabilistic forecast skill. Figures~\ref{fig:healda-vs-era5-full} and \ref{fig:healda-vs-ifs-full} report additional metrics, including RMSE, anomaly correlation coefficient (ACC), ensemble spread, and the spread--skill ratio (SSR), along with the deterministic skill (skill of individual ensemble members).

Across variables, the deterministic skill of HealDA-initialized FCN3 is similar to IFS ENS and ERA5-initialized FCN3 member skill. In contrast, all ensemble metrics (CRPS, ensemble RMSE, and ensemble ACC) show a larger gap, with HealDA-initialized forecasts typically trailing ERA5/IFS by less than one day. The spread of HealDA- and ERA5-initialized FCN3 forecasts is effectively identical, again indicating FCN3 is largely unaffected by the distribution shift of HealDA analyses and its forecast dynamics remain unchanged. Compared to IFS ENS, HealDA-initialized forecasts have a significantly lower SSR for much of the forecast due to HealDA producing a deterministic initial condition not capturing the uncertainty, as opposed to IFS ENS members each having initial conditions generated from perturbing a deterministic control member \citep{ecmwf2023ifs_ens}.

\subsection{Effect of Initial Condition Resolution}
\label{sec:ic-resolution}

\begin{figure}
    \centering
    \includegraphics[width=1.0\linewidth]{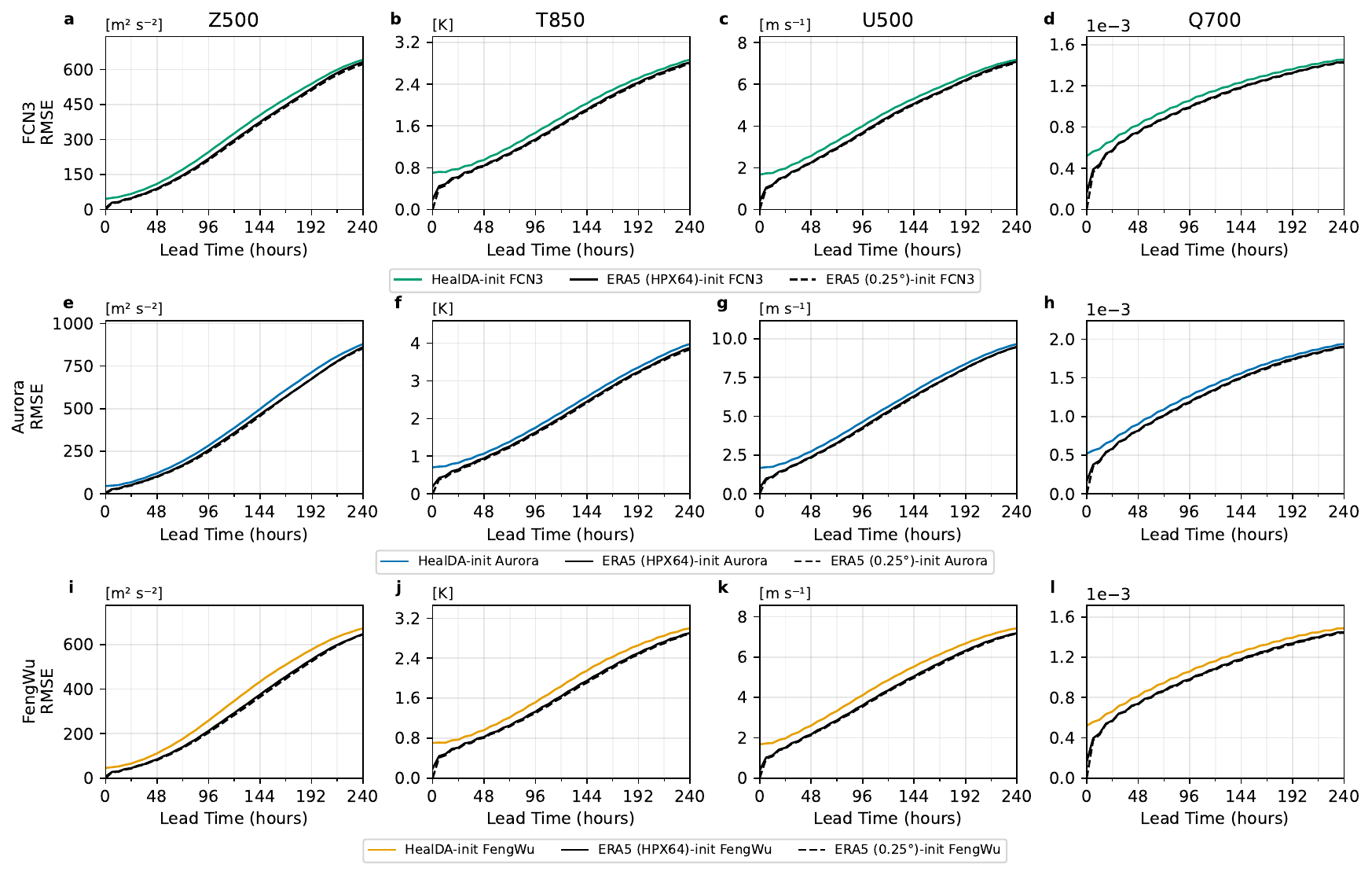}
    \caption{
    \textbf{Forecast skill from $0.25^\circ$ and $1^\circ$ initial conditions.}
    RMSE skill across FCN3 (a--d), Aurora (e--h), and FengWu (i--l) forecasting models
    when initialized from HealDA, ERA5 HPX64, and ERA5 $0.25^\circ$ initial conditions.
    Scores are averaged over initializations in 2022 at 06/18 UTC.}
    \label{fig:ic-resolution-effect}
\end{figure}

All results shown so far initialize the forecast models from $1^\circ$ HPX64 analyses, even though the underlying models are trained at $0.25^\circ$. Figure~\ref{fig:ic-resolution-effect} shows that initializing FCN3, Aurora, and FengWu with $0.25^\circ$ ERA5 inputs yields only a small skill improvement—equivalent to less than $\sim$3~h of lead time—which is smaller than the skill gap between HealDA and ERA5 HPX64 initializations. These $0.25^\circ$-initialized runs are evaluated using the same scoring protocol as before (i.e., all fields are scored on the HPX64 grid). This supports the view that large-scale accuracy dominates error growth, while small-scale differences in the initial condition are of secondary importance. Even relative to ERA5 $0.25^\circ$ initialization, HealDA-initialized forecasts lose less than $<$24~h of effective lead time.

\begin{figure}
    \centering
    \includegraphics[width=1.0\linewidth]{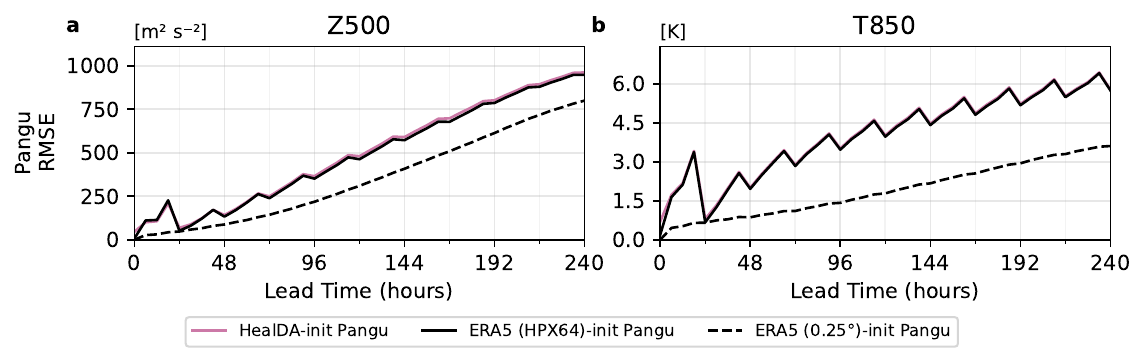}
    \caption{
    \textbf{Pangu-Weather forecasts}
    RMSE skill of Pangu-Weather when initialized from HealDA, ERA5 HPX64, and ERA5 $0.25^\circ$ initial conditions. Scores are averaged over initial conditions in 2022 at the 06/18 UTC.}
    \label{fig:pangu}
\end{figure}

We further try initializing Pangu-Weather, another $0.25^\circ$ forecast model, with HealDA initial conditions, shown in Figure \ref{fig:pangu}. We use the combined 6-hour and 24-hour Pangu forecast models in the original configuration, using the 24-hour lead time model at day multiples and the 6-hour model elsewhere. For the first three 6-hour forecast steps, Pangu exhibits a period of significant instability for both ERA5 HPX64 and HealDA initial conditions, likely reflecting that its Swin Transformer architecture \citep{liu2021swintransformerhierarchicalvision} is sensitive to the input distribution from the lower-resolution HPX64 inputs. This instability is lessened after the first 24-hour model timestep; however, the overall skill is quite poor compared to using $0.25^\circ$ ERA5 initial conditions. As such, we exclude Pangu from other comparisons. We note that Pangu-Weather is the only model we find to be sensitive to the input resolution.

\subsection{Power Spectra of HealDA-initialized Forecasts}
\label{sec:spectra}

\begin{figure}
    \centering
    \includegraphics[width=1.0\linewidth]{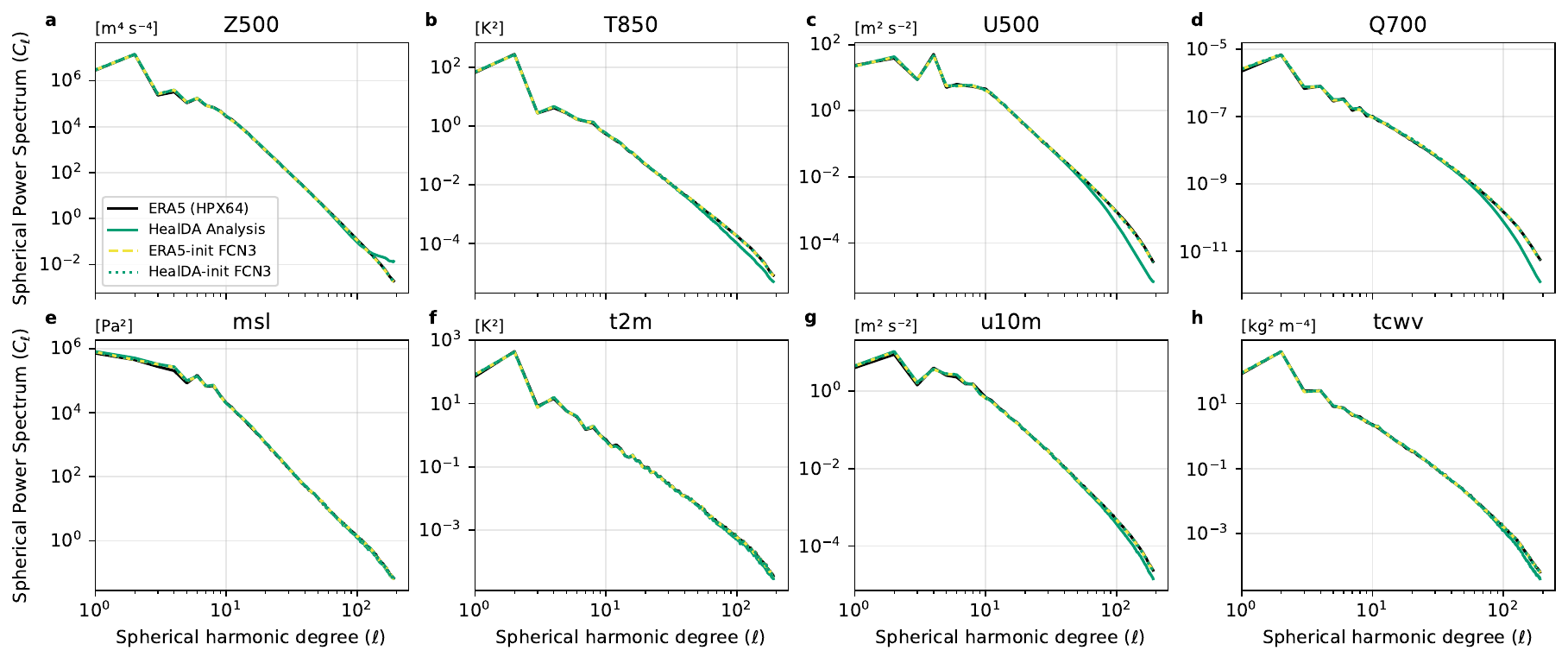}
    \caption{
    \textbf{Spectra of HealDA-initialized FCN3 forecasts.}
    Average spherical power spectra on the HPX64 grid for ERA5, HealDA analysis, and FCN3 forecasts initialized from HealDA and ERA5. Forecast spectra are averaged over forecasts and lead times in 2022, and the analysis spectra are averaged over the 2022 period.}
    \label{fig:spectra}
\end{figure}
To assess how HealDA's small-scale smoothing interacts with the forecast model, we examine spherical power spectra of FCN3 forecasts initialized from HealDA across lead times in Fig.~\ref{fig:spectra}. At $t=0$, HealDA analyses exhibit noticeably lower variance at high $\ell$ than ERA5 for almost all variables, with the exception of Z500, where we see increased small-scale noise relative to the target ERA5 at HPX64, largely owing to the unique larger dynamic range of geopotential fields. The overall trend is consistent with the small-scale blurring expected from a regression-trained DA network.

However, as the forecast evolves, this initial difference disappears: the lead-time-averaged spectra from both initializations lie almost exactly on top of each other and match the target ERA5 HPX64 spectrum. In other words, FCN3's learned dynamics regenerate the small-scale variance missing from the HealDA analysis and are largely insensitive to the modest distribution shift between ERA5 and HealDA initial conditions. This supports the conclusion from the forecast skill curves that the additional error in HealDA-initialized runs is driven primarily by increased large-scale error in the initial condition, rather than by any loss of effective resolution.

\subsection{Analysis Error Bias}
\label{sec:analysis-bias}
\begin{figure}[t]
  \centering
  \includegraphics[width=1.0\linewidth]{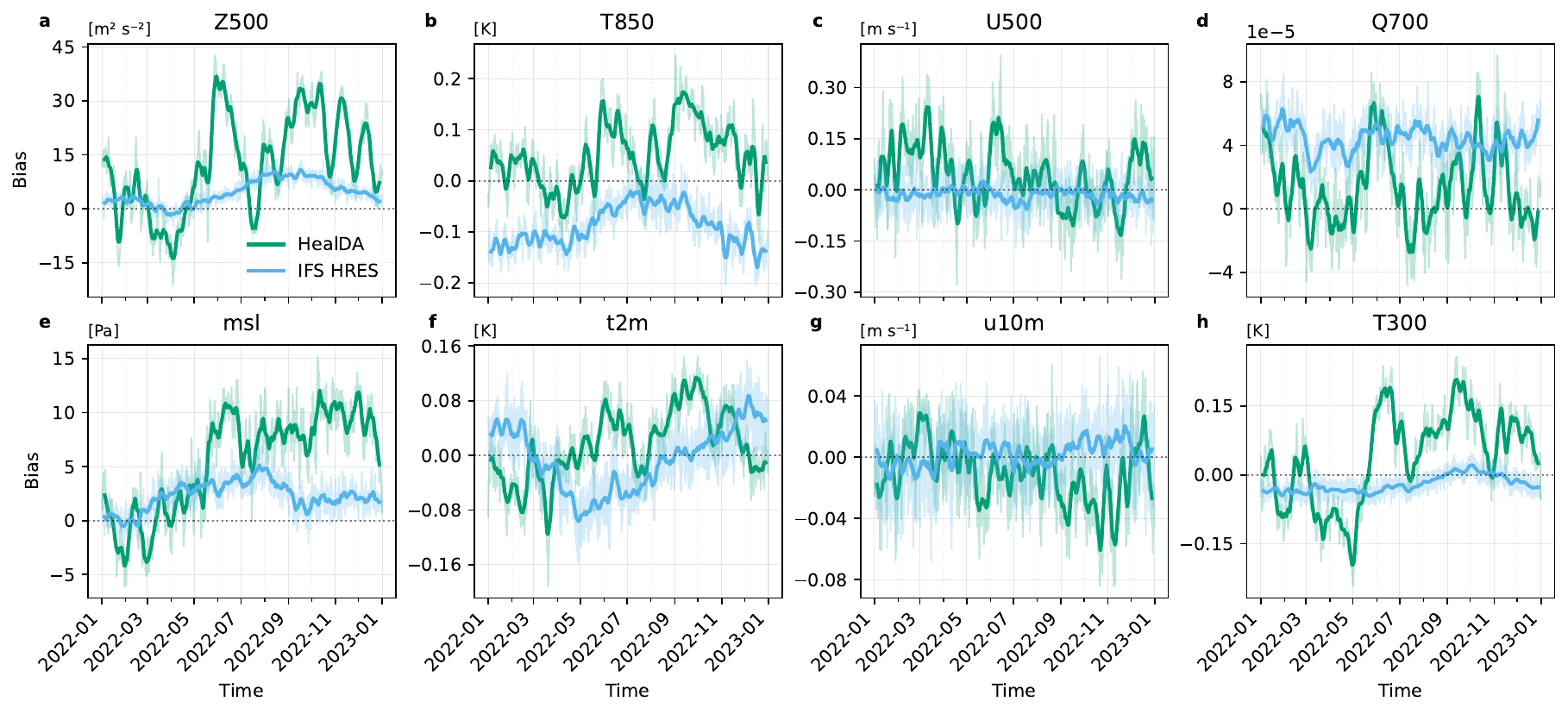}
  \caption{
    \textbf{Mean Bias Error of the HealDA Analysis} Time series of the mean bias error for both HealDA and IFS against ERA5 in the 2022 test period, computed every 6 hours (00/06/12/18 UTC). The original data is shown with reduced opacity to reduce noise, and the solid line represents the 7-day moving average.}
  \label{fig:bias}
\end{figure}

Figure \ref{fig:bias} shows a 1-year time series of mean bias error for HealDA and IFS HRES relative to ERA5. Across most fields, the bias of both DA systems is generally zero-centered, with HealDA exhibiting substantially more temporal variability. For geopotential and mean sea level pressure fields, there is a slight positive bias towards the end of 2022. Some of these periods of larger bias magnitude correspond to those in which our observation dataset has dropouts (see \ref{sec:robustness}). Expanding the assimilated sensor suite (e.g., adding infrared sounders) would better constrain the analysis and likely reduce this variability.

\subsection{Choice of Initialization Time}
\label{sec:init-time}
\begin{figure}[t]
  \centering
  \includegraphics[width=1.0\linewidth]{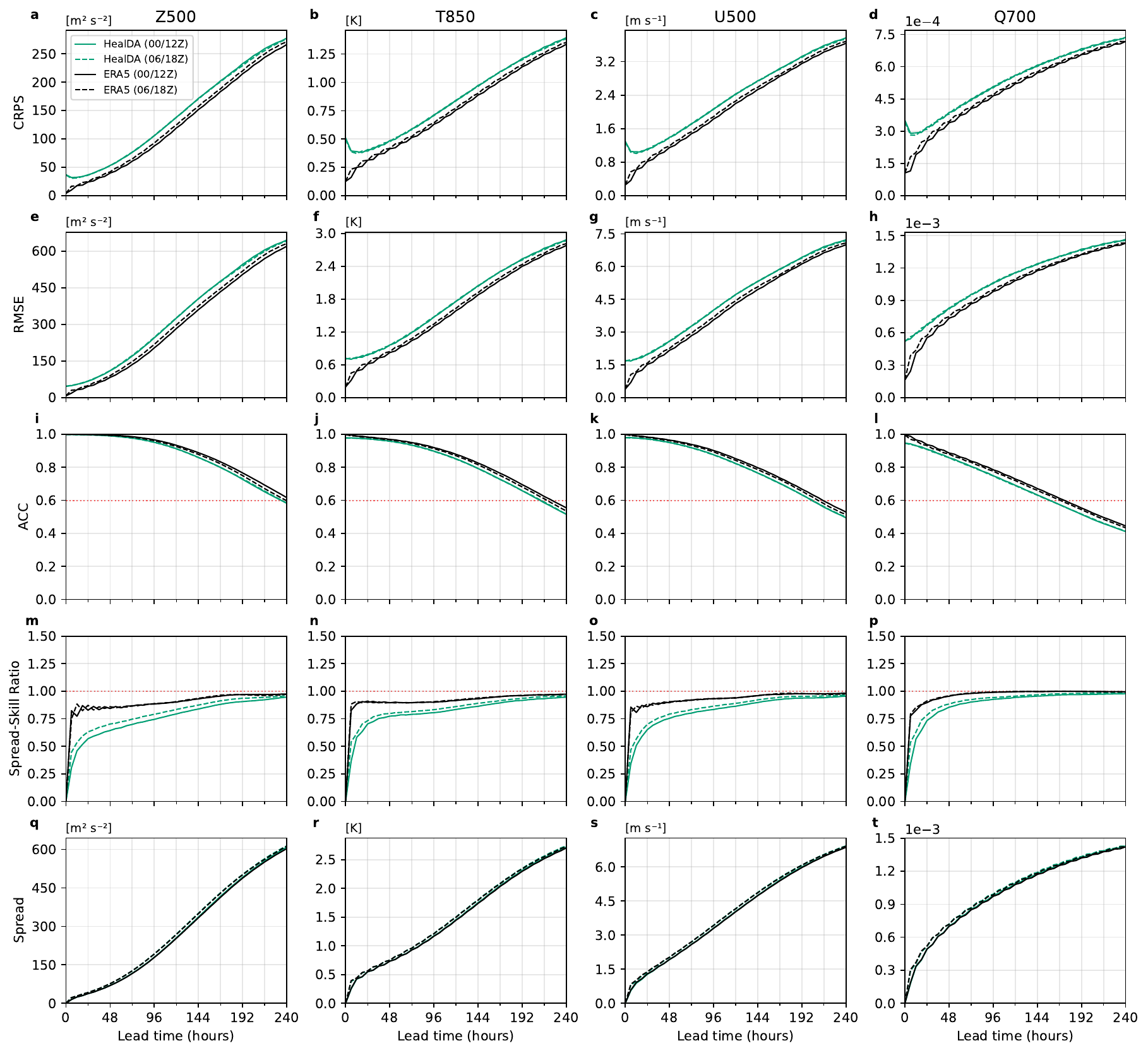}
  \caption{
    \textbf{Effect of init time on FCN3 forecast skill.} Forecast skill of HealDA- and ERA5-initialized FCN3 forecasts at the 00/12 (solid line) and 06/18 (dashed line) UTC initialization times. Scores are computed against ERA5 at the HPX64 resolution across 128 times for each set of initialization times. Red dotted lines mark reference thresholds (ACC = 0.6; SSR = 1).}
  \label{fig:init-effect}
\end{figure}

Figure \ref{fig:init-effect} shows the effect of the init times on FCN3 forecasts. For ERA5 initializations, 00/12 UTC init forecasts perform better than 06/18 UTC inits, giving on the order of a 3-6 hour benefit in lead time. This is because ERA5 has a 9-hour lookahead at 00/12 UTC as opposed to a 3-hour lookahead at 06/18 UTC. FCN3, having trained on ERA5 at both the 00/12 UTC and 06/18 UTC, implicitly learns this difference and produces forecasts with higher spread and marginally lower skill when initialized at 06/18 UTC. Our HealDA-initialized FCN3 forecasts are largely unaffected, since HealDA here is inferenced with a 3-hour lookahead. Given this discrepancy between the ERA5 and HealDA assimilation windows, unless specified otherwise, we always use 06/18 UTC inits to compare ERA5- and HealDA-initialized forecasts for the fairest comparison.

\subsection{Choice of Observation Window}
\label{sec:obs-window}
\begin{figure}[t]
  \centering
  \includegraphics[width=0.5\linewidth]{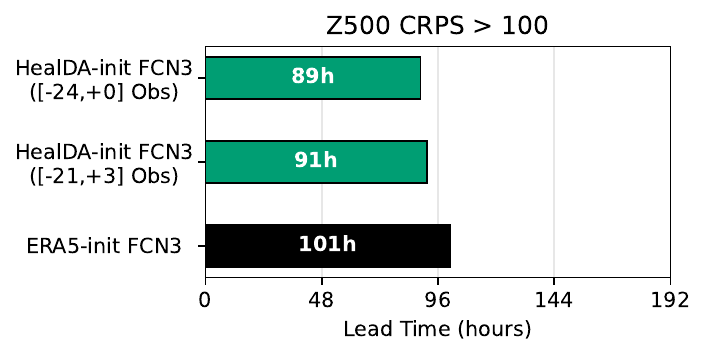}
    \caption{\textbf{Z500 CRPS threshold-crossing time for different HealDA observation windows.} Bars show the lead time (hours) at which the Z500 CRPS first exceeds 100 (linearly interpolated and rounded to the nearest hour) for FCN3 forecasts initialized from HealDA analyses that assimilated observations in $[-24,0]$ or $[-21,+3]$\,h relative to the analysis time. ERA5-initialized FCN3 is shown as a reference. Longer times indicate better forecast skill. Scores computed against ERA5 across 06/18 UTC initializations.}
  \label{fig:obs-window}
\end{figure}

To assess the sensitivity of HealDA to the observation lookahead, we evaluate the $[-21,+3]$\,h trained checkpoint using two different observation windows in \ref{fig:obs-window}: $[-24,0]$ and $[-21,+3]$\,h relative to the analysis time. Removing the $+3$\,h lookahead results in a nearly uniform degradation of CRPS throughout the forecast, corresponding to an approximately $\sim$3\,h shift in the skill curves (i.e., similar skill is reached about 3\,h earlier) and an average $\sim$3\% increase in CRPS across lead times. This behavior reflects a simple temporal shift in the effective analysis time rather than a structural change in the analysis quality.

This flexible inference is possible because each observation is explicitly time-stamped. Thus, the same trained model can be inferenced with different observation windows without retraining, enabling flexible operation in real time using all available observations, or in a delayed mode that mirrors the typical latency of operational analyses.

\subsection{ACC Skill Summary}
\begin{figure}
    \centering
    \includegraphics[width=1\linewidth]{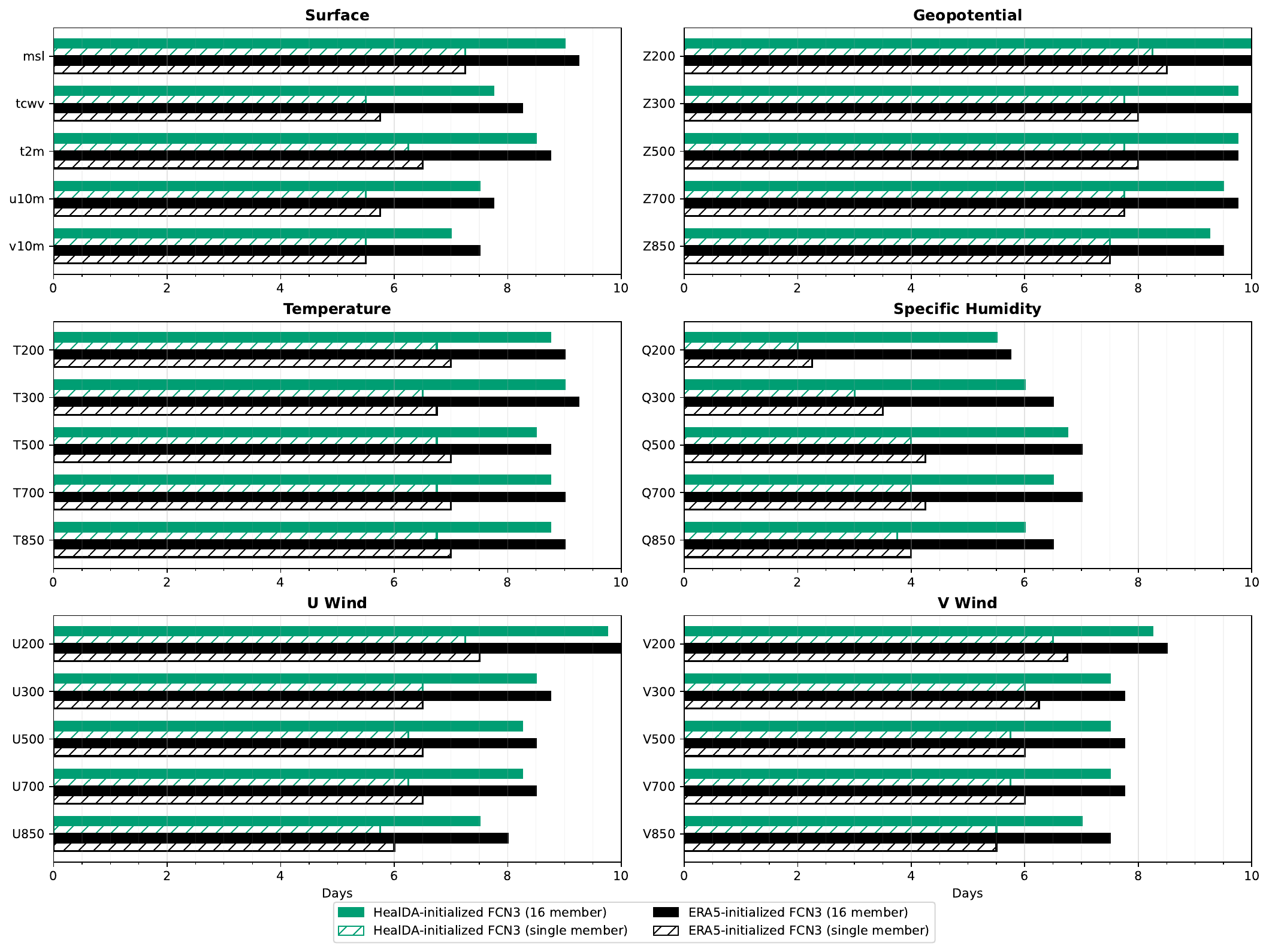}
    \caption{Comparisons of skillful forecast lead times of FCN3 forecasts with DA and ERA5 inits. Skillful forecast lead time is defined as the Anomaly Correlation Coefficient (ACC) being above 0.6. The scores are averaged over 128 initial conditions at 06/18 UTC in 2022.}
    \label{fig:acc-summary}
\end{figure}

Figure \ref{fig:acc-summary} summarizes the skillful lead times of the ERA5- and HealDA-initialized FCN3 forecasts. The HealDA forecasts have only a 6 to 12-hour lower skillful lead time than ERA5-initialized forecasts. This clearly demonstrates that our HealDA analyses are not far behind ERA5 in terms of information content and forecasting skill.

\subsection{Impact of Level 2 Products}
\label{sec:lvl2-ablation}

The original model’s input observation set included some Level~2 products, namely satellite-derived wind retrievals and GNSS-RO temperature and humidity channels. These products were initially included because they appeared to have a limited impact on overall skill, were available operationally, and were not expected to introduce strong dependence on upstream background information. However, we later found that the GNSS-RO temperature and humidity channels (\texttt{Temperature\_\allowbreak at\_\allowbreak Obs\_\allowbreak Location}, \texttt{Specific\_\allowbreak Humidity\_\allowbreak at\_\allowbreak Obs\_\allowbreak Location}) correspond to fields in the GSI diagnostic pipeline obtained by interpolating the background state to the observation location, rather than quantities derived from the GNSS-RO measurements. Given our focus on a direct observation-to-state formulation without a background state as an input field, we investigate the impact of removing all Level~2 products here.

To do so, we consider two ablation settings without Level~2 products. In the first, we remove all Level~2 products only at inference time from the original checkpoint, referred to as HealDA (no L2 at inference). Since Level~2 products constitute a large fraction of the conventional observation stream used by the original model, removing them only at inference time is expected to have an outsized negative impact on analysis quality relative to their true importance. We therefore also fine-tune the original HealDA checkpoint using only Level~1 inputs, referred to as HealDA (L1), to achieve the best possible skill with Level~1 products alone. This second setting allows the model to adapt to the absence of Level~2 products and provides a more meaningful estimate of their importance to the analysis.

\begin{figure}
    \centering
    \includegraphics[width=1\linewidth]{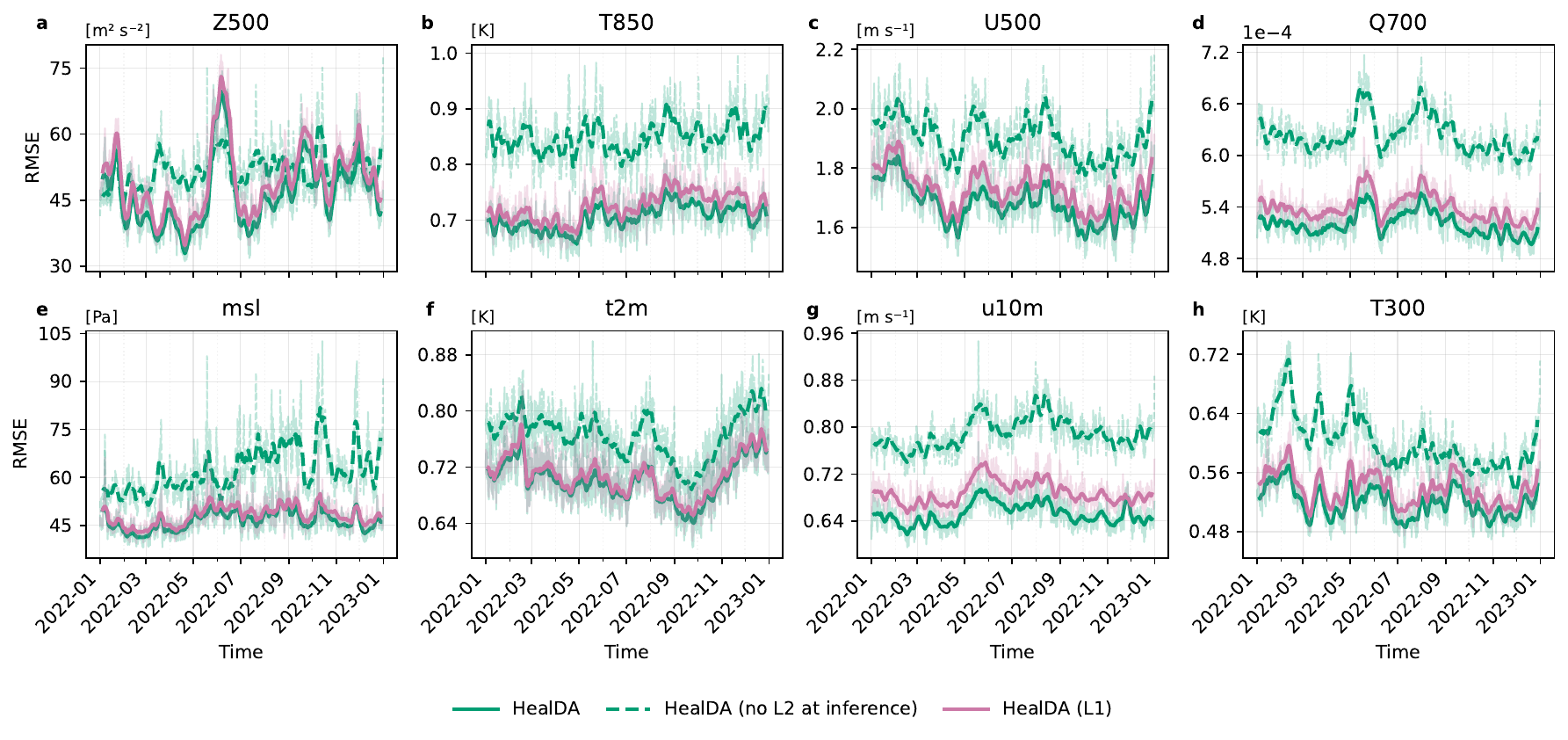}
    \caption{
    \textbf{Impact of removing Level~2 products on HealDA analysis RMSE.}
    Time series of global RMSE for HealDA, HealDA (no L2 at inference), and HealDA (L1), all scored against ERA5 over the 2022 test period. HealDA (no L2 at inference) denotes removing Level~2 products only at inference time from the original checkpoint, while HealDA (L1) denotes the variant fine-tuned using only Level~1 inputs. The original data are shown with reduced opacity to reduce noise, and the solid lines represent 7-day moving averages.
    }
    \label{fig:analysis-rmse-nolvl2}
\end{figure}

\begin{figure}
    \centering
    \includegraphics[width=1\linewidth]{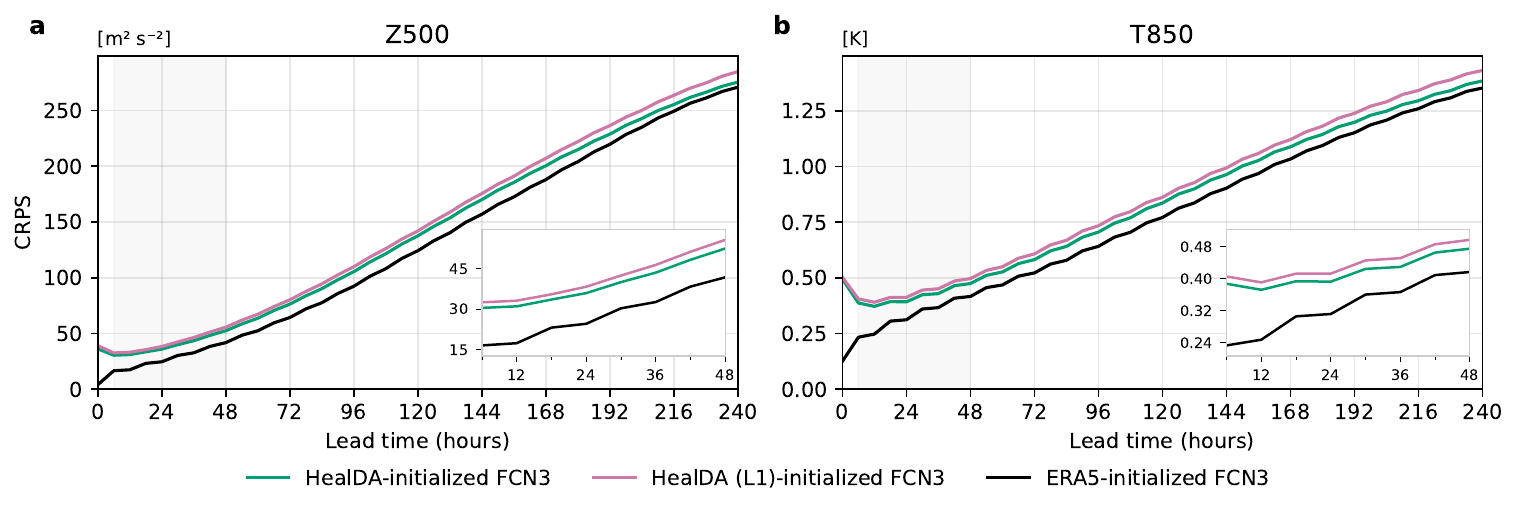}
    \caption{
    \textbf{Forecast impact from removing Level~2 products.}
    CRPS skill curves of 16-member FCN3 forecasts initialized from ERA5, HealDA, and HealDA (L1) (the variant fine-tuned using only Level~1 inputs) analyses. Forecasts are verified against ERA5 on the HPX64 grid and averaged over 128 initial conditions at 06/18 UTC in 2022. The inset panels zoom into the 6--48\,h lead-time range.
    }
    \label{fig:crps-nolvl2}
\end{figure}

\begin{figure}
    \centering
    \includegraphics[width=1\linewidth]{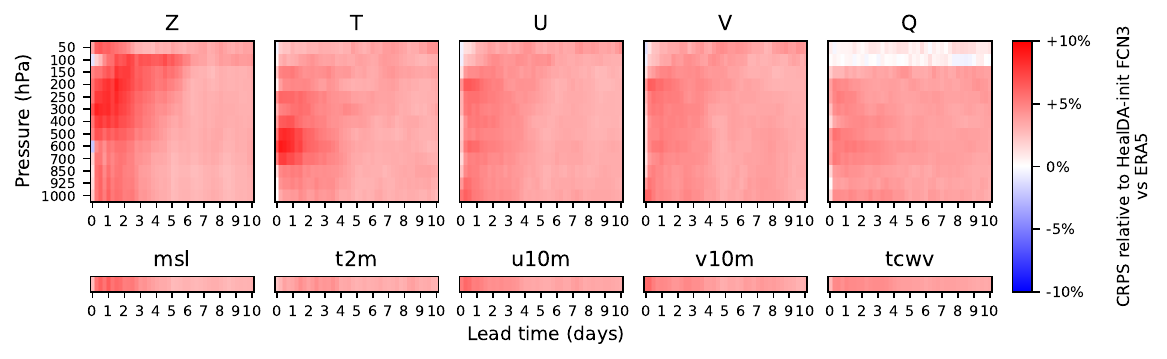}
    \caption{
    \textbf{Scorecard of HealDA (L1)-initialized forecasts relative to HealDA.}
    Relative CRPS differences between FCN3 forecasts initialized from HealDA (L1) analyses (the variant fine-tuned using only Level~1 inputs) and from HealDA analyses. Forecasts are initialized at 06/18 UTC and verified against ERA5 in 2022. Positive values indicate increased CRPS relative to HealDA-initialized forecasts, while negative values indicate improved CRPS.
    }
    \label{fig:scorecard-nolvl2}
\end{figure}

For HealDA (L1), we initialize the model from the original released HealDA weights and continue training for approximately $10\%$ of the original training duration using only Level~1 inputs. We use a linear warmup for 50{,}000 samples to one-tenth of the original peak learning rate, followed by cosine decay to zero over 1M samples. Figure~\ref{fig:analysis-rmse-nolvl2} compares analysis RMSE over the 2022 test year for HealDA, HealDA (no L2 at inference), and HealDA (L1).

Simply removing Level~2 products at inference time increases analysis error by $10$--$20\%$ for most fields, confirming that the released checkpoint relies strongly on those inputs. For example, Z500 RMSE increases from $46.6~\mathrm{m^2\,s^{-2}}$ to $51.6~\mathrm{m^2\,s^{-2}}$ (\(+10.9\%\)). The wind fields show some of the clearest degradation near the surface, with the largest increases in u10m, U1000, and U925, consistent with scatterometer winds acting as a strong near-surface ocean-wind constraint \citep{ecmwf-seminar}. By contrast, fine-tuning without Level~2 products recovers much of the lost skill. Relative to the original model, HealDA (L1) increases Z500 RMSE to $49.3~\mathrm{m^2\,s^{-2}}$ (\(+5.9\%\)), with similar \(3\)--\(6\%\) increases in analysis error for other fields. Thus, much of the degradation from naive inference-time removal reflects distribution shift, rather than an irreducible dependence on Level~2 products.

We next assess the downstream forecast impact of removing Level~2 observations in the assimilation. Specifically, we initialize FCN3 from the original HealDA analyses and from HealDA (L1) analyses, and evaluate CRPS against ERA5 over the 2022 test year. Relative to the original HealDA analyses, HealDA (L1) analyses yield \(3\)--\(6\%\) increase in CRPS across the majority of variables and lead times, summarized in Figure~\ref{fig:scorecard-nolvl2}. The CRPS skill curves in Figure~\ref{fig:crps-nolvl2} illustrate that the removal of Level~2 products leads to about \(3\)--\(6\) hours of effective forecast lead-time loss.

Nevertheless, the downstream forecasting conclusion remains unchanged. In the main text, HealDA-initialized FCN3 forecasts were found to trail ERA5-initialized FCN3 by less than one day of effective lead time. After fine-tuning without Level~2 products, this gap increases slightly: for example, at 06/18 UTC initializations, the lead time at which Z500 CRPS first exceeds 100 lags the ERA5-initialized forecast by 13 hours, compared with 10 hours previously. Thus, removing Level~2 products degrades both the initial conditions and the downstream forecasts, but does not alter the main result that the analyses produced by the observation-only HealDA framework remain usable as plug-and-play initial conditions for off-the-shelf forecast models.

\begin{table}[t]
\centering
\setlength{\belowcaptionskip}{6pt}
\caption{\textbf{Variables predicted by HealDA.} Atmospheric variables are predicted at 13 pressure levels: 1000, 925, 850, 700, 600, 500, 400, 300, 250, 200, 150, 100, and 50~hPa. All targets are from ERA5 reanalysis.}
\label{tab:healda-outputs}
\small 
\begin{tabular}{lll}
\toprule
\textbf{Type} & \textbf{Variable} & \textbf{Abbreviation} \\
\midrule
\textbf{Atmospheric} 
    & Geopotential & Z \\
    & Temperature & T \\
    & Zonal wind & U \\
    & Meridional wind & V \\
    & Specific humidity & Q \\
\midrule
\textbf{Surface} 
    & 2 meter temperature & t2m \\
    & 10 meter $u$-wind component & u10m \\
    & 10 meter $v$-wind component & v10m \\
    & 100 meter $u$-wind component & u100m \\
    & 100 meter $v$-wind component & v100m \\
    & Mean sea-level pressure & msl \\
    & Total column water vapor & tcwv \\
    & Sea surface temperature & sst \\
    & Sea ice concentration & sic \\
\bottomrule
\end{tabular}
\end{table}

\begin{table}[h!]
\centering
\caption{HealDA Model and Observation Encoder Hyperparameters}
\label{tab:hyperparams}
\begin{tabular}{ll}
\hline
\textbf{Hyperparameter} & \textbf{Value} \\
\hline
\multicolumn{2}{c}{\textbf{Training}} \\
Learning rate (HPX ViT backbone) & $5\times10^{-4}$ \\
Learning rate (Obs encoder) & $1\times10^{-4}$ \\
Learning rate schedule & Linear warm-up (50k), cosine decay (10M) \\
Optimizer & AdamW ($\beta_1=0.9,\ \beta_2=0.95$) \\
Weight decay & 0.05 \\
Dropout probability & 0.05 \\
Drop path probability & Linear schedule (0 $\rightarrow$ 0.1 across layers) \\
Batch size & 8 \\
Training duration & $\sim$8.3 days (1600 GPU hours) \\
Epochs & $\sim$333 \\
Loss function & Huber loss ($\delta=0.1$) \\
Hardware & 1 H100 node (8 GPUs) \\
Model size & DiT-L, 330M params, 24 blocks, dim=1024 \\
ViT sequence length & 12{,}288 (HPX32 / HPX Level 5) \\
\hline
\multicolumn{2}{c}{\textbf{Observation Encoder}} \\
number of sensors & 5 \\
number of channels per sensor & 5-22\\
$K$ (token dimension) & 32 \\
$D_{\text{meta}}$ (metadata dimension) & 28 \\
$D$ (feature dimension) & 512 \\
Obs Tokenizer MLP & 2 layers, SiLU activation \\
Metadata encoding & Fourier features (time, height, pressure) \\
\hline
\end{tabular}

\end{table}

\end{document}